\documentclass{jfm}
\usepackage{graphicx}
\usepackage{bm}
\usepackage{epstopdf, epsfig}
\usepackage[dvipsnames]{xcolor}
\usepackage{hyperref}
\hypersetup{colorlinks, citecolor = BlueViolet, filecolor = blue, linkcolor = blue, urlcolor = blue}
\usepackage{makecell}

\shorttitle{Thermal boundary layer structure in low-$Pr$ convection}
\shortauthor{Ambrish Pandey}

\title{Thermal boundary layer structure in low-Prandtl-number turbulent convection}

\author{Ambrish Pandey
  \corresp{\email{ambrish.pandey@nyu.edu}}
  }

\affiliation{Center for Space Science, New York University Abu Dhabi, PO Box 129188, Abu Dhabi, UAE
}

\date{\today}

\newcommand{\mc}{\mathcal}

\begin{document}

\maketitle

\begin{abstract}

We study the structure of the thermal boundary layer (BL) in Rayleigh-B\'enard convection for Prandtl number ($Pr$) 0.021 by conducting direct numerical simulations in a two-dimensional square box for Rayleigh numbers ($Ra$) up to $10^9$. The large-scale circulation in the flow divides the horizontal plates into three distinct regions, and we observe that the local thermal BL thicknesses in the plume-ejection region are larger than those in the plume-impact and shear-dominated regions. Moreover, the local BL width decreases as $Ra^{-\beta(x)}$, with $\beta(x)$ depending on the position at the plate. We find that $\beta(x)$ are nearly the same in impact and shear regions and are smaller than those in the ejection region. Thus, the local BL width decreases faster in the ejection region than those in the shear and impact regions, and we estimate that the thermal BL structure would be uniform throughout the horizontal plate for $Ra \geq 8 \times 10^{12}$ in our low-$Pr$ convection. We compare the thermal BL profiles measured at various positions at the plate with the Prandtl-Blasius-Pohlhausen (PBP) profile and find deviations everywhere for all the Rayleigh numbers. However, the dynamically-rescaled profiles, as suggested by Zhou \& Xia ({\it Phys. Rev. Lett.}, vol. 104, 2010, 104301), agree well with the PBP profile in the shear and impact regions for all the Rayleigh numbers, whereas they still deviate in the ejection region. We also find that, despite the growing fluctuations with increasing $Ra$, thermal boundary layers in our low-$Pr$ convection are transitional and not yet fully turbulent. 

\end{abstract}


\section{Introduction} \label{sec:intro}

Turbulent flows driven by thermal convection occur commonly in nature. For example, the flows in the convection zone of the Sun and the Earth's outer core are driven primarily by the buoyancy force arising due to the inhomogeneous temperature field~\citep{Schumacher:RMP2020, Hanasoge:ARFM2016, Pandey:Nature2018}. The Prandtl number ($Pr$), which is the ratio of the kinematic viscosity $\nu$ and the thermal diffusivity $\kappa$ of a fluid, is approximately $10^{-6}$ in the solar convection zone~\citep{Schumacher:RMP2020} and $Pr \approx 10^{-2}$ in the Earth's outer core~\citep{Schumacher:PNAS2015}. Rayleigh-B\'enard convection (RBC) is a paradigm of turbulent convection flows in nature, where a fluid kept between two horizontal plates is heated from below and cooled from above~\citep{Ahlers:RMP2009, Chilla:EPJE2012, Verma:NJP2017, Verma:book2018}. The main governing parameters of RBC are the Prandtl and Rayleigh numbers, where $Ra$ indicates the strength of the thermal driving force compared to the viscous dissipative forces in the flow. Thin viscous and thermal boundary layers (BLs) near the isothermal horizontal plates exist in RBC, and the behaviour of the flow in the BL region remains laminar-like even up to very large $Ra$ despite a highly turbulent flow in the bulk region away from the walls. Properties of a low-$Pr$ convection flow differ in certain aspects than those of the high-$Pr$ flows.  In low-$Pr$ RBC, the thermal BL is thicker compared to the viscous BL, and therefore, directly interacts with the turbulent bulk flow. Moreover, low-$Pr$ RBC is dominated by the inertial effects and is highly turbulent compared to high-$Pr$ convection flows at the same $Ra$~\citep{Schumacher:PNAS2015, Pandey:POF2016, Scheel:PRF2017, Shishkina:PRF2017, Pandey:Nature2018}. Structure of the thermal BL has primarily been explored for moderate~\citep{Shi:JFM2012, Wagner:JFM2012, Scheel:JFM2014} and high-$Pr$ RBC~\citep{Werne:PRE1993, Lui:PRE1998, Wang:EPJB2003, Zhou:POF2011}, where the thermal BL is either of a similar width as the viscous BL or nested within the latter. In this paper, we study the horizontal structure of the thermal BL in a low-$Pr$ RBC. 

Characterization of the thermal BL is important as it controls the global heat transport, which is quantified using the Nusselt number ($Nu$)~\citep{Grossmann:JFM2000}. In RBC, the mean thermal BL width can be computed as $H/(2 Nu)$~\citep{Ahlers:RMP2009, Chilla:EPJE2012}, and this relation has been verified for a wide range of $Ra$ and $Pr$~\citep{Stevens:JFM2011, Scheel:JFM2012, Scheel:JFM2014, Scheel:JFM2016, Scheel:PRF2017, Schumacher:PRF2016, Zhou:JFM2013, Zhang:PRE2017, Bhattacharya:POF2019a}. This relation arises from the definition of the thermal BL thickness using the slope method~\citep{Wagner:JFM2012} and due to the fact that the heat transport is purely diffusive at the horizontal plates. The thermal BL thickness, however, is a local quantity and varies in magnitude at the horizontal plates. For high-$Pr$  RBC, local BL thickness has been observed to be the smallest near the center of the plate and increases symmetrically~\citep{Lui:PRE1998} or asymmetrically~\citep{Werne:PRE1993, Wang:EPJB2003, Zhou:POF2011} in the plane of the large-scale circulation (LSC) as the sidewalls are approached. For moderate-$Pr$ RBC, \citet{Wagner:JFM2012} observed that the local BL thickness increases almost linearly along the direction of LSC, whereas \citet{Scheel:JFM2014} observed that the local BL thicknesses are larger at the plume-detachment locations. In this paper, we find that the local BL thickness in low-$Pr$ RBC varies asymmetrically along the plate, and its relative variation in the central region of the plate decreases from around 3.1 at $Ra = 5 \times 10^5$ to around 1.6 at $Ra = 10^9$. 

In turbulent RBC, nearly all the imposed temperature difference occurs primarily in the thin thermal BLs, whereas the bulk region remains mostly isothermal. The thermal BL profiles in RBC have been compared with the PBP profile, which was originally proposed for a laminar shear flow on a semi-infinite heated plate~\citep{Landau:FM1987, Shishkina:NJP2010}, and systematic deviations with increasing $Ra$ and decreasing $Pr$ have been reported~\citep{Shishkina:JFM2009, Shishkina:PRL2015, Shishkina:PRF2017, Shi:JFM2012, Stevens:PRE2012, Zhou:JFM2010, Zhou:POF2011, Ovsyannikov:EJMB2016, Wang:PRF2016, Wang:JFM2018}.  As the Prandtl-Blasius BL theory is based on the two-dimensional (2D) equations, the BL profiles obtained from 2D flows are more likely to be closer to the PBP profile. \cite{Poel:JFM2013} compared the thermal BL profiles at the center of the horizontal plate in two- and three-dimensional (3D) RBC for $Pr = 4.38$ and $Ra \approx 10^8$, and observed that the BL profile in 2D is indeed closer to the PBP profile compared to that in 3D. They credited the larger deviation from the PBP profile in 3D RBC to an increased plume activity compared to that in 2D RBC. The agreement with the PBP profile has been observed to improve if the profiles are averaged in a dynamical frame of reference based on the instantaneous BL thicknesses~\citep{Zhou:JFM2010, Zhou:POF2011, Scheel:JFM2012, Stevens:PRE2012, Shi:JFM2012}. Nonetheless, persistent deviations even after using this dynamic rescaling have been reported in 3D RBC for moderate and high-$Pr$ RBC~\citep{Shi:JFM2012, Scheel:JFM2012, Stevens:PRE2012}. The local thermal BL profiles have not been compared with the PBP profile in low-$Pr$ RBC, except for the horizontally-averaged  profiles, which exhibit increasing deviation with decreasing $Pr$~\citep{Scheel:JFM2016, Shishkina:PRF2017, Ching:PRR2019}. Therefore, we measure the temperature profiles at various horizontal positions in our low-$Pr$ RBC and observe deviations from the PBP profile everywhere, with the degree of deviation depending on the measurement position. 

As mentioned above, the properties of the near-wall temperature field have mostly been studied in moderate- and high-$Pr$ RBC as the investigations of low-$Pr$ convection are inhibited due to several experimental and numerical challenges. On the one hand, the opaqueness of low-$Pr$ fluids, such as mercury, gallium, or liquid sodium, restricts the use of the optical measurement techniques~\citep{Cioni:JFM1997, Glazier:NATURE1999, Zuerner:JFM2019}. On the other hand, numerical investigations of low-$Pr$ convection require massive computational resources as very small length and time scales need to be resolved accurately to prudently study them~\citep{Schumacher:PRF2016,  Scheel:JFM2016,  Scheel:PRF2017, Pandey:POF2016, Shishkina:PRF2017, Pandey:Nature2018}.  Note that exploring convection in very high-$Pr$ fluids also requires significant computational resources as the smallest length scale in the temperature field, the Batchelor scale $\eta_B$, becomes much finer compared to the Kolmogorov length scale $\eta_K$, which is the smallest length scale in the velocity field~\citep{Silano:JFM2010, Horn:JFM2013, Poel:JFM2013, Pandey:PRE2014, Shishkina:PRL2015, Shishkina:PRF2017}. These two length scales are related as $\eta_B = \eta_K/\sqrt{Pr}$~\citep{Shishkina:NJP2010}, and thus, it is the Batchelor scale that needs to be resolved properly to prudently study high-$Pr$ convection. Due to growing computational resources,  properties of the temperature field in the BL region have been explored  only recently in low-$Pr$ RBC~\citep{Scheel:JFM2016, Schumacher:PRF2016, Scheel:PRF2017, Shishkina:PRF2017}.

\citet{Scheel:JFM2016} computed the local thermal BL thicknesses in a cylindrical cell of aspect ratio unity for $Pr = 0.005, 0.021, 0.7$ using the local vertical temperature gradient at the horizontal plates and observed that the mean BL thicknesses are larger when the regions near the sidewall are included. They also observed that the horizontally- and temporally-averaged thermal BL profile for the lower Prandtl numbers deviates from the corresponding profile for $Pr = 0.7$. \citet{ Scheel:PRF2017} computed the displacement thicknesses and shape factors (defined respectively in equations~(\ref{eq:dis_thick}) and~(\ref{eq:shape}) here) of the mean temperature profiles and found that the displacement thicknesses increase with decreasing $Pr$ and the shape factors deviate from those of the corresponding PBP profiles.  \citet{Schumacher:PRF2016} and \citet{Scheel:PRF2017} compared the mean temperature profiles in low-$Pr$ RBC with the fully turbulent BL profile, which exhibits a logarithmic region, and observed that the region where the log-scaling is observed in the profiles increases with increasing $Ra$.  \citet{Shishkina:PRF2017} proposed an analytical form of the mean thermal BL profile by incorporating the effects of turbulent fluctuations in the BL equations, and observed very good agreement with their numerically computed profiles in a cylindrical RBC cell of aspect ratio unity for $Pr$ between 0.01 and 2547.9. The detailed horizontal structure of the thermal BL in low-$Pr$ convection is, however, still unexplored and is the primary objective of this paper.

In this work, we conduct direct numerical simulations (DNS) of RBC in a low-$Pr$ fluid at high Rayleigh numbers and study the structure of the thermal BL. We simulate convection flows for $Pr = 0.021$, which is a typical Prandtl number of mercury or gallium, and for $Ra = 5 \times 10^5 - 10^9$ in a 2D square box. Although the turbulent flows in nature are three-dimensional, the characteristics of some of them can be understood using the two- or quasi-two-dimensional models. For instance, turbulent convective flows under the effects of a strong rotation or a strong magnetic field behave similarly to a quasi-2D flow~\citep{Chandrasekhar:book}. In RBC of aspect ratio around unity, the LSC is usually the strongest flow structure, and the BL structure has primarily been  explored along the direction of LSC~\citep{Lui:PRE1998, Wang:EPJB2003, Wagner:JFM2012}. However, the plane of LSC does not remain fixed in 3D RBC, which poses an additional challenge in the study of the BL structure as one has to be in the direction of LSC at every instant. Furthermore, we choose a 2D geometry as (i) almost all the BL theories have been developed for two-dimensional flows, (ii) the measurement probes always remain in the plane of LSC, in contrast to RBC in a cylindrical cell, where the plane of LSC exhibits reorientation~\citep{Wagner:JFM2012, Schumacher:PRF2016, Zuerner:JFM2019}, (iii) and high Rayleigh numbers can be achieved even with moderate computational resources; the highest $Ra$ explored in this work has not been achieved in 3D DNS at this $Pr$~\citep{Scheel:PRF2017}. Note that 2D RBC has been utilized to better understand some of the important phenomena in convection, e.g., the properties of flow reversals~\citep{Sugiyama:PRL2010, Chandra:PRL2013, Podvin:JFM2015, Pandey:PRE2018, Zhang:JFM2020}, the onset of the ultimate regime of convection~\citep{Zhu:PRL2018}, the logarithmic temperature profiles~\citep{Poel:PRL2015, Zhu:PRL2018} to name a few. We detect LSC in our simulations, which yields three different regions, namely, the plume-ejection, shear-dominated, and plume-impact regions, at the horizontal plates~\citep{Poel:PRL2015, Schumacher:PRF2016, Zhu:PRL2018}. Using our DNS data, we explore the horizontal dependence of the local thermal BL thickness and find that the local thicknesses in the ejection region are larger compared to those in the shear and impact regions. We measure the temperature profiles in the aforementioned regions and observe that they deviate from the PBP profile for all $Ra$. However, once the profiles in the shear and impact regions are dynamically rescaled~\citep{Zhou:PRL2010}, they agree very well with the PBP profile. We also find that due to growing turbulent fluctuations with increasing $Ra$, the local temperature profiles in the ejection region become increasingly similar to the fully turbulent thermal BL profile.  However, the explored Rayleigh numbers are still not large enough to yield a fully turbulent thermal BL.

\section{Details of direct numerical simulations} \label{sec:sim}

Conservation of momentum, internal energy, and mass lead to equations that govern the dynamics of RBC~\citep{Chilla:EPJE2012, Verma:NJP2017, Verma:book2018}. The nondimensional governing equations under the Oberbeck-Boussinesq approximations are
\begin{eqnarray}
\frac{\p \boldsymbol{u}}{\p t} + \boldsymbol{u} \bcdot \bnabla \boldsymbol{u} & = & -\bnabla p + T \hat{\boldsymbol{z}} + \sqrt{\frac{Pr}{Ra}}   \nabla^2 \boldsymbol{u}, \label{eq:u} \\ 
\frac{\p T}{\p t} + \boldsymbol{u} \bcdot \bnabla T & = &  \frac{1}{\sqrt{Ra Pr}} \nabla^2 T, \label{eq:T} \\ 
\bnabla \bcdot \boldsymbol{u} & = & 0, \label{eq:m}
\end{eqnarray}
where $\boldsymbol{u} = (u_x,u_z), T$, and $p$ are respectively the velocity, temperature, and pressure fields defined on a two-dimensional bounded domain. The above equations are nondimensionalized using $H, \Delta T, u_f$, and $t_f$ as the length, temperature, velocity, and time scales respectively, where $u_f = \sqrt{\alpha g \Delta T H}$ is the free-fall velocity and $t_f = H/u_f$ is the free-fall time. The Rayleigh number is defined as $Ra = \alpha g \Delta T H^3/\nu \kappa$,  where $\alpha$ is the thermal expansion coefficient of the working fluid, $g$ is the acceleration due to gravity, and $\Delta T$ is the temperature difference between the top and bottom plates separated by distance $H$. 

We perform direct numerical simulations in a two-dimensional square box of length $L = H = 1$ by integrating equations~(\ref{eq:u}--\ref{eq:m}). We employ no-slip condition for the velocity field on all the boundaries. The horizontal plates are isothermal and the sidewalls are adiabatic. We use a spectral element solver {\sc Nek5000}~\citep{Fischer:JCP1997, Scheel:NJP2013} to simulate the RBC flow for $Pr = 0.021$ for $Ra = 5 \times 10^5 - 10^9$. The flow domain is divided into $N_e$ spectral elements, and the turbulence fields are expanded within each element using $N^\mathrm{th}$-order Legendre polynomials. Thus, we probe our flow domain with $N_e N^2$ mesh cells. We use a denser grid near all the boundaries to capture the strong variations of the velocity and temperature fields in the BLs. Important parameters of our simulations are also summarized in table~\ref{table:sim_detail}.

\begin{table}
\begin{center}
\begin{tabular}{lccccccccccccc}
Run & $Ra$ & $N_e$ & $N$ & $N_\mathrm{bl}$ & $Nu \pm \Delta Nu$ & $Nu_{\varepsilon_T}$ & $Nu_{\varepsilon_u}$ & $Re \pm \Delta Re$ & $\frac{\Delta_\mathrm{max}}{\eta_K}$ & \thead{$t_\mathrm{total}$ \\ $(t_f)$} & $N_s$ \\
1 & $5 \times 10^5$ & $5184$ & 7 & 180 & $5.2 \pm 0.01$ & 5.2 & 5.2 & $6012 \pm 1$ & 2.8 &  121 & 575 \\
1a & $5 \times 10^5$ & $52900$ & 3 & 100 & $5.2 \pm 0.01$ & 5.2 & 5.2 & $6011 \pm 1$ & 0.58 & 114 & 125 \\
2 & $10^6$ & $5184$ & 7 & 170 & $6.3 \pm 0.05$ & 6.3 & 6.3  & $6806 \pm 13$ & 3.5 &  111 & 400 \\
2a & $10^6$ & 52900 & 3 & 90 & $6.3 \pm 0.01$ & 6.3 & 6.3  & $6775  \pm 82$ & 0.74 & 133 & 250 \\
3 & $10^7$ & $52900$ & 9 & 210 & $10.8 \pm 0.01$ & 10.8 & 10.7 & $22473 \pm 410$  & 0.56 & 101 & 340 \\
3a & $10^7$ & $52900$ & 5 & 120 & $11.1 \pm 0.5$ & 11.1 & 11.0 & $22187 \pm 110$  & 0.98 & 132 & 400 \\
4 & $10^8$ & $52900$ & 11 & 215 & $19.6 \pm 0.1$ & 19.7  & 19.3 & $97118 \pm 1566$ & 0.97 & 97 & 590 \\
5 & $10^9$ & $198916$ & 13 & 360 &  $35.4 \pm 1.6$ & 36.6 & 33.9 & $432930 \pm 191$ & 0.85 &  8.6 & 452 \\
5a & $10^9$ & $198916$ & 7 & 190 &  $36.2 \pm 0.1$ & 37.4 & 34.7 & $433305 \pm 420$ & 1.5 & 14.4 & 717 \\
\end{tabular}
\caption{Important parameters of our DNS runs at a fixed $Pr = 0.021$ in a 2D unit square box. Here, $N_e$ is the total number of spectral elements in the flow domain, $N$ is the order of Legendre interpolation polynomials within each element, $N_\mathrm{bl}$ is the number of grid points within each thermal BL, $Nu, Nu_{\varepsilon_T}$, and $Nu_{\varepsilon_u}$ are the globally- and temporally-averaged Nusselt numbers computed using equations~(\ref{eq:Nu}), (\ref{eq:Nu_epst}), and (\ref{eq:Nu_epsv}), respectively, $Re$ is the Reynolds number computed using the root-mean-square (rms) velocity, $\Delta_{\mathrm{max}}/\eta_K$ is the ratio of the maximum grid spacing to the Kolmogorov length scale in the flow, $t_\mathrm{total}$ is the total integration time  after reaching the statistically steady state in units of the free-fall time $t_f$, and $N_s$ is total number of equidistant snapshots stored for the duration $t_\mathrm{total}$ and used for the analyses in this paper. The errorbars in $Nu$ and $Re$ indicate the difference between the mean values computed over the first and second halves of the datasets.}
\label{table:sim_detail}
\end{center}
\end{table}

We start our simulations from the conduction state with random perturbations and wait until a statistically steady state is reached, i.e., when the time-averaged values of the global quantities, such as the convective heat flux and total kinetic energy, do not change significantly. 
For instance, we separately compute $Nu$ and $Re$ for the first and second halves of the datasets, and denote the difference between the mean values over the two halves as $\Delta Nu$ and $\Delta Re$, respectively~\citep{Scheel:JFM2014, Scheel:JFM2016}. In table~\ref{table:sim_detail}, we list $\Delta Nu$ and $\Delta Re$ as the errorbars in $Nu$ and $Re$, which show that the mean values of the global quantities  in our simulations do not vary by more than 5\% in the steady state. Due to a larger time scale for the momentum diffusion compared to that for the heat diffusion, the smallest structures in the velocity field in low-$Pr$ convection are much finer compared to the smallest structures in the temperature field~\citep{Scheel:JFM2016, Pandey:POF2016}. Therefore, it is crucial to adequately resolve very fine spatial and temporal Kolmogorov scales in low-$Pr$ RBC. The Kolmogorov length scale is computed as $\eta_K = (\nu^3/\varepsilon_u)^{1/4}$, where $\varepsilon_u$ is the viscous dissipation rate defined as
\begin{equation}
\varepsilon_u({\bm x}) = \frac{\nu}{2} \left( \frac{\partial u_i}{\partial x_j} + \frac{\partial u_j}{\partial x_i} \right)^2. \label{eq:epsv}
\end{equation}
Here, $u_i$ is the component of the velocity in the $x_i$-direction. For a well-resolved simulation, the maximum grid spacing $\Delta_\mathrm{max}$ in the entire flow domain should be smaller than or comparable to $\eta_K$ and $\eta_B$. The Batchelor scale in our low-$Pr$ flow is coarser compared to the Kolmogorov scale. Therefore, we estimate the Kolmogorov scale using the globally and temporally-averaged viscous dissipation rate and list the ratio $\Delta_\mathrm{max}/\eta_K$ in table~\ref{table:sim_detail}.  We can see that for most of our simulations $\Delta_\mathrm{max}/\eta_K$ is smaller than one, which indicates that the smallest length scales in our flow are resolved adequately.

An important quantity in RBC is the Nusselt number, which is defined as the ratio of the total heat transport to that occurred by conduction alone~\citep{Ahlers:RMP2009, Chilla:EPJE2012, Verma:book2018}. The globally- and temporally-averaged Nusselt number in our nondimensional units is computed as 
\begin{equation}
Nu = 1 + \sqrt{Ra Pr} \, \langle u_z T \rangle_{A,t}, \label{eq:Nu}
\end{equation}
where $\langle \cdot \rangle_{A,t}$ denotes the averaging over the entire simulation domain and the integration time. The Nusselt number can also be computed using the exact relations in RBC as~\citep{Shraiman:PRA1990, Zhang:JFM2017}
\begin{eqnarray}
Nu_{\varepsilon_u} & = & 1 + \frac{H^4}{\nu^3} \frac{Pr^2}{Ra} \langle \varepsilon_u \rangle_{A,t}, \label{eq:Nu_epsv} \\
Nu_{\varepsilon_T} & = & \frac{H^2}{\kappa (\Delta T)^2} \langle \varepsilon_T \rangle_{A,t}. \label{eq:Nu_epst}
\end{eqnarray}
Here $\varepsilon_T$ is the thermal dissipation rate defined as the rate of loss of thermal energy per unit mass and computed as
\begin{equation}
\varepsilon_T(\bm x) = \kappa \left[ \left( \frac{\partial T}{\partial x} \right)^2 + \left( \frac{\partial T}{\partial z} \right)^2 \right]. \label{eq:epst}
\end{equation}

The requirement of resolving very fine Kolmogorov scales significantly increases the computational effort to explore convection at low Prandtl numbers~\citep{Poel:JFM2013, Schumacher:PNAS2015, Schumacher:PRF2016, Pandey:POF2016, Scheel:JFM2016, Scheel:PRF2017, Pandey:Nature2018, Zwirner:JFM2020}. Due to inadequate spatial resolution, the velocity and temperature derivatives, and in turn, the viscous and thermal dissipation rates, are inaccurately estimated. In our spectral element simulations, this inadequacy of spatial resolution is reflected in the vertical profiles of the dissipation rates, which do not vary smoothly at the element boundaries~\citep{Scheel:NJP2013}. Moreover, the Nusselt numbers obtained from the dissipation rates using equations~(\ref{eq:Nu_epsv}--\ref{eq:Nu_epst}) differ from $Nu$ computed using equation~(\ref{eq:Nu}). Therefore, the adequacy of the spatial resolution can also be ensured by comparing $Nu$ computed using the aforementioned three methods. In table~\ref{table:sim_detail}, we list $Nu_{\varepsilon_u}$ and $Nu_{\varepsilon_T}$ along with $Nu$ for all the simulations, and find that they agree reasonably well; the largest difference appears for $Ra = 10^9$, which is due to a limited statistics for this simulation. Furthermore, due to strong variations of the velocity and temperature fields within the BLs, the number of mesh cells should be sufficient in these regions~\citep{Shishkina:NJP2010}. Table~\ref{table:sim_detail} shows that the number of grid points within the thermal BL  is huge for all the Rayleigh numbers, thus indicating that the thermal BLs are resolved very well in our simulations.  In addition, we check the grid-sensitivity by performing simulations for $Ra = 5 \times 10^5, 10^6, 10^7$, and $10^9$ with different spatial resolutions, and find that the integral quantities, such as the Nusselt and Reynolds numbers, as well as the BL structure, remain nearly the same, which also indicates that our flows are properly resolved. 

\section{Global transports and flow structure} \label{sec:global}

\subsection{Global quantities}

As we study low-$Pr$ convection in a two-dimensional domain, we first compare the scaling of the global quantities in our simulations with those observed in three-dimensional RBC for $Pr \approx 0.021$. 

The Reynolds number $Re$ is another important global quantity in RBC, which is a measure of the turbulent momentum transport in the flow. We compute $Re$ in our flow as
\begin{equation}
Re = \sqrt{Ra/Pr} \, u_\mathrm{rms} \, , \label{eq:Re}
\end{equation}
with the rms velocity $u_\mathrm{rms}$ defined as
\begin{equation}
u_\mathrm{rms} = \sqrt{ \langle u_x^2+u_z^2 \rangle_{A,t}} \, .
\end{equation}
Important theoretical models  in RBC yielding the scaling relations for the global quantities, such as $Nu$ and $Re$,~\citep{Shraiman:PRA1990, Grossmann:JFM2000} assume the existence of a large-scale structure of the order of the size of the cell. For instance, the \citet{Shraiman:PRA1990} theory provides the scalings of $Nu$ and $Re$ by considering the properties of (2D) viscous and thermal BLs generated due to the shear applied by the LSC.  Similarly, the Grossmann-Lohse model~\citep{Grossmann:JFM2000} is based on the existence of an LSC in the convection cell,  which is characterized by a single velocity scale. Thus, the above scaling theories are applicable in both 2D and 3D RBC~\citep{Poel:JFM2013, Zhu:PRL2018}. As the LSC also exists in our low-$Pr$ 2D RBC for all the Rayleigh numbers, it is interesting to compare the global quantities in our flow with those observed in 3D RBC for similar governing parameters, where the LSC structure is also observed. 

\begin{figure}
\centerline{
\includegraphics[width=\textwidth]{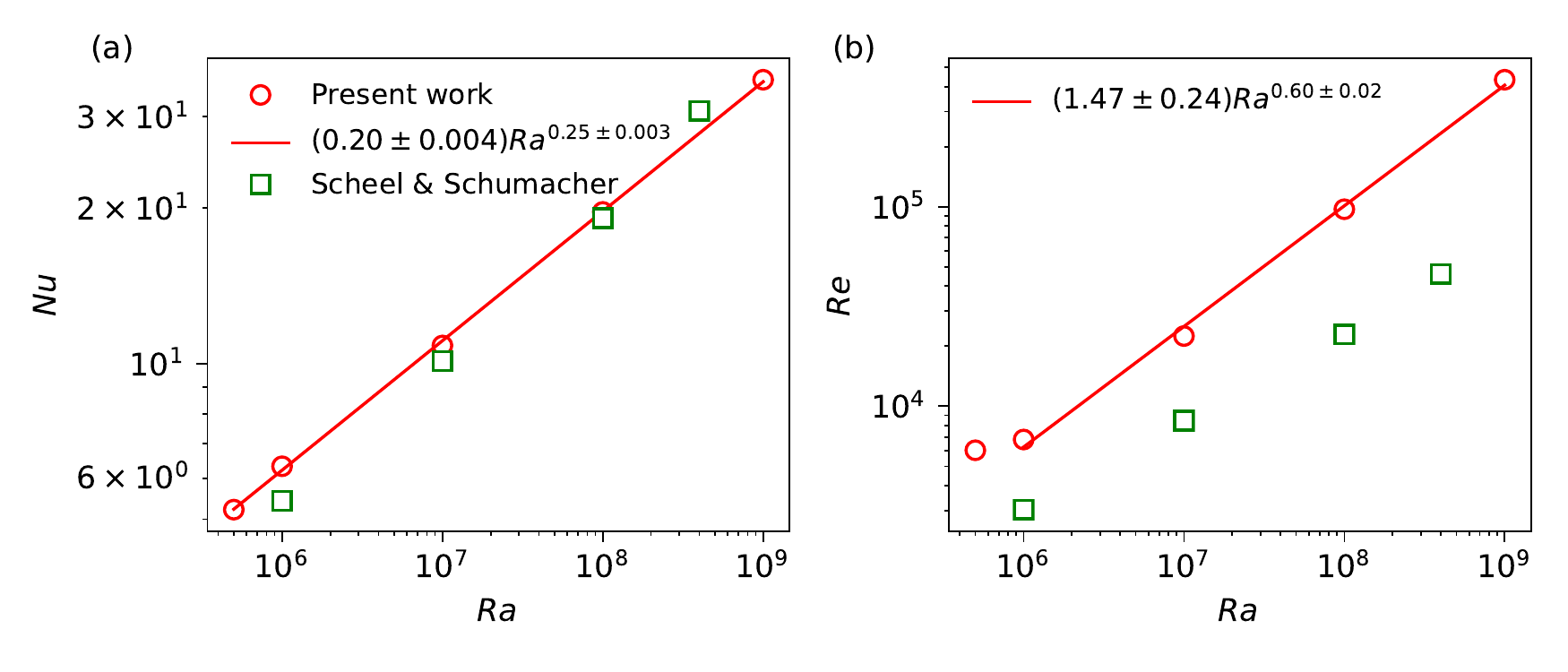}
}
\caption{The Nusselt and Reynolds numbers as a function of $Ra$ in our 2D RBC (circles) and those obtained by~\citet{Scheel:PRF2017} for $Pr = 0.021$ in a cylindrical cell of aspect ratio unity (squares). A reasonably good agreement of $Nu$ in both 2D and 3D RBC for $Pr = 0.021$ is remarkable.}
\label{fig:nu_re}
\end{figure}

We compute the Nusselt and Reynolds numbers using equations~(\ref{eq:Nu}) and~(\ref{eq:Re}) for all our simulation runs and plot them as a function of $Ra$ in figure~\ref{fig:nu_re}. To compare our $Nu$ and $Re$ with those obtained in a 3D RBC, we also show $Nu$ and $Re$ for $Pr = 0.021$ from \citet{Scheel:PRF2017}, who investigated low-$Pr$ RBC in a cylindrical cell of aspect ratio one from $Ra = 3 \times 10^5$ to $Ra = 4 \times 10^8$. Thus their governing parameters are very similar to the parameters in the present study, and due to the existence of a similar large-scale structure in both these flows, it is interesting to compare the global heat and momentum transports between these flows. Figure~\ref{fig:nu_re}(a) reveals that $Nu$ in our 2D RBC increases as $Ra^\gamma$ and agree reasonably well with those from \citet{Scheel:PRF2017}. The best fit to our data yields $Nu = (0.20 \pm 0.004)Ra^{0.25 \pm 0.003}$, which is close to $Nu = (0.13 \pm 0.04)Ra^{0.27 \pm 0.01}$ observed by \citet{Scheel:PRF2017} for nearly the same range of $Ra$. 
The exponent $\gamma = 0.25$ in our flow is smaller than $\gamma \approx 0.30$ observed for $Pr = 0.7$ and 5.3 in 2D RBC with a similar flow configuration~\citep{Zhang:JFM2017}. The lower $\gamma$ in our low-$Pr$ 2D RBC is similar to that observed in 3D RBC, where $\gamma$ also decreases with decreasing $Pr$~\citep{Scheel:PRF2017}. Note that the Grossmann-Lohse theory~\citep{Grossmann:JFM2000} also predicts a lower $\gamma$ for convection in low-$Pr$ fluids compared to those in moderate and high-$Pr$ fluids.

We would like to point out that even the magnitudes of $Nu$ obtained in our 2D flows are very similar to those obtained by \citet{Scheel:PRF2017}. This is remarkable and indicates that, at least in the present range of parameters, the LSC is probably the most dominant mode of heat transport in both the 2D and 3D RBC. \citet{Poel:JFM2013} also compared $Nu$ between 2D and 3D RBC with similar flow configurations and noticed that $Nu$ at $Ra = 10^8$ are closer in 2D and 3D for low Prandtl numbers, whereas a stronger deviation was observed at moderate Prandtl numbers.  In an earlier study, \citet{Schmalzl:EPL2004} observed, however, that the integral quantities in 2D and 3D RBC differ for low Prandtl numbers, whereas are similar for high Prandtl numbers~\citep{Poel:JFM2013, Pandey:Pramana2016}. However, the sizes of the flow domains in 2D and 3D cases were different, in addition to a smaller $Ra = 10^6$ used in~\citet{Schmalzl:EPL2004}. Interestingly, the scaling $Nu \sim Ra^{0.25}$ in our 2D RBC agrees well with the existing literature for $Pr \approx 0.021$ in 3D RBC  as $\gamma \approx 0.25$ has been observed in various earlier investigations~\citep{Cioni:JFM1997, Glazier:NATURE1999, Grossmann:JFM2000, Pandey:POF2016, Zuerner:JFM2019}. Note, however, that the structure and the characteristics of LSC are more complex in 3D RBC, where it has a quasi-2D character and exhibits twisting and sloshing modes in addition to azimuthal reorientations in a cylindrical cell~\citep{Wagner:JFM2012, Schumacher:PRF2016, Zwirner:JFM2020}, which are clearly absent in 2D RBC. Therefore, the observed similarity of $Nu-Ra$ scaling in our low-$Pr$ convection with those in 3D RBC requires a more detailed investigation, which is beyond the scope of the present work.

Figure~\ref{fig:nu_re}(b) exhibits the Reynolds number in our simulations as a function of $Ra$ along with the $Re$ for $Pr = 0.021$ from~\citet{Scheel:PRF2017}. It is clear that the Reynolds numbers in our 2D RBC are consistently higher compared to those in the 3D RBC for all $Ra$. The best fit in the range of $Ra$ from $10^6$ to $10^9$ yields $Re = (1.47 \pm 0.24)Ra^{0.60 \pm 0.02}$, which is different from $Re \sim Ra^{0.45}$ observed in 3D RBC for $Pr \approx 0.021$~\citep{Scheel:PRF2017, Pandey:POF2016}. Note that the magnitude of the Reynolds numbers as well as the exponent in the $Re-Ra$ scaling in 2D RBC have been observed to be higher compared to those in 3D RBC for moderate and high Prandtl numbers  too~\citep{Poel:JFM2013, Zhang:JFM2017}.  Thus, the exponent in the $Re-Ra$ scaling in 2D RBC appears to remain nearly the same ($\approx 0.6$) for a wide range of Prandtl numbers~\citep{Poel:JFM2013, Pandey:Pramana2016, Zhang:JFM2017}. The larger magnitude of $Re$ in our flow is probably due to the more coherent motion of the thermal plumes in 2D RBC than in 3D RBC~\citep{Zhang:PRE2017}.

To summarize, the scaling of $Nu$ in our 2D RBC, which is related to the scaling of the mean thermal BL thickness, is very similar to that observed in 3D RBC for $Pr \approx 0.021$. Therefore, the characteristics of the thermal BL in our 2D convection may also be similar to those in 3D RBC for low Prandtl numbers.

\subsection{Flow structure}

\begin{figure}
\centerline{
\includegraphics[width=0.8\textwidth]{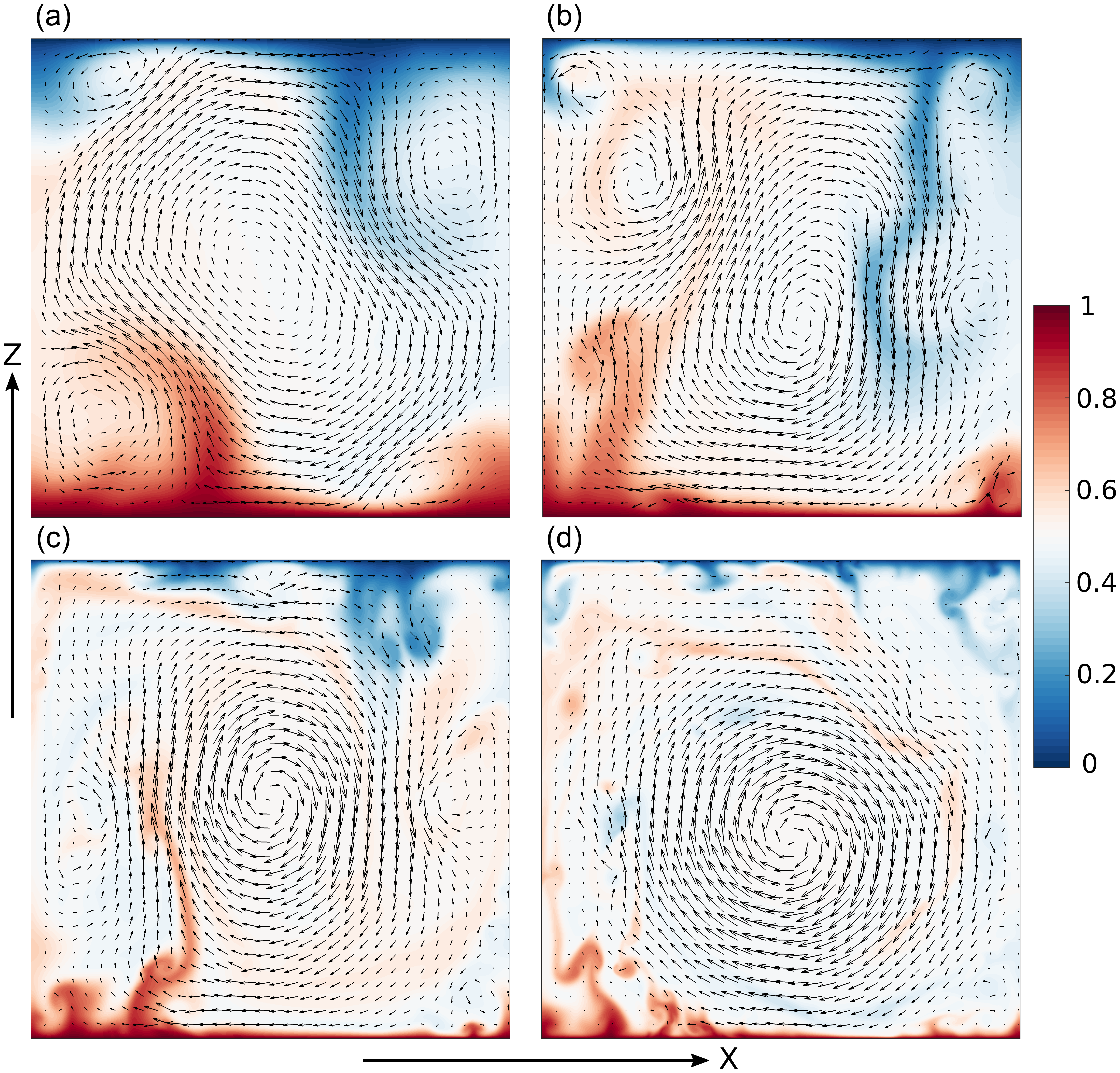}
}
\caption{Instantaneous temperature (colors) and velocity (vector) fields in the entire simulation domain for (a) $Ra = 10^6$, (b) $Ra = 10^7$, (c) $Ra = 10^8$, and (d) $Ra = 10^9$. Increasingly finer thermal structures are observed in the flow  with increasing $Ra$. Also, the primary flow structure i.e., the LSC, becomes stronger, whereas the secondary structures become weaker with the increasing Rayleigh number.}
\label{fig:flow_inst}
\end{figure}

\begin{figure}
\centerline{
\includegraphics[width=0.8\textwidth]{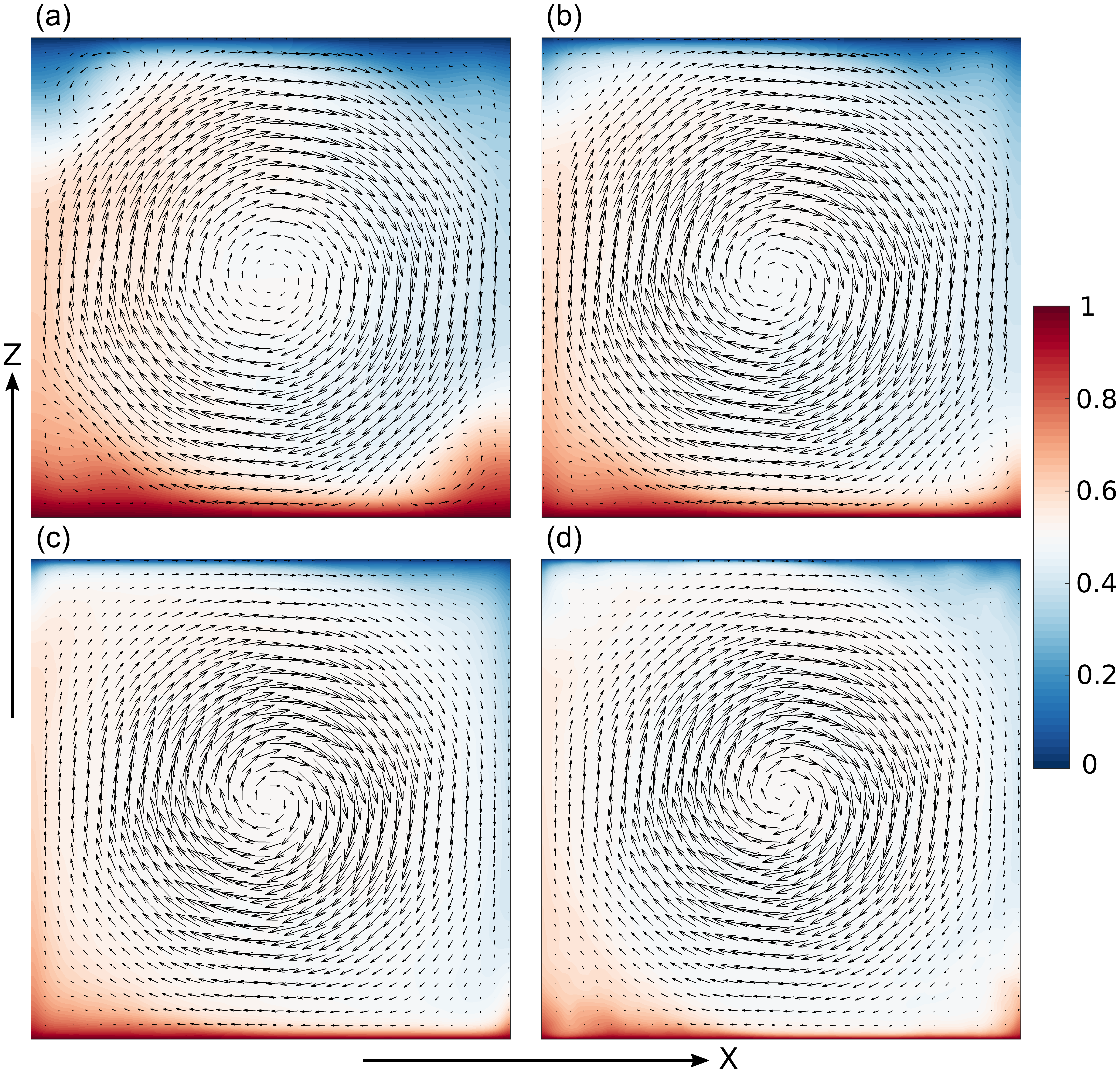}
}
\caption{Time-averaged flow structure for (a) $Ra = 10^6$, (b) $Ra = 10^7$, (c) $Ra = 10^8$, and (d) $Ra = 10^9$. The temperature (colors) and the velocity (vector) fields averaged for the entire duration of simulations exhibit that the LSC structure gets increasingly developed, i.e., occupies a larger fraction of the flow domain, with increasing $Ra$.}
\label{fig:flow_mean}
\end{figure}
 
The thermal plumes emitted from the top and bottom plates in an RBC cell of aspect ratio around unity move coherently by forming an LSC, which we also detect in our simulations.  For all the Rayleigh numbers, we observe that the LSC rotates in a single direction for the entire duration of the simulation, and thus, does not exhibit any flow reversal~\citep{Sugiyama:PRL2010, Chandra:PRL2013, Podvin:JFM2015, Pandey:PRE2018, Zhang:JFM2020}. Specifically, the LSC rotates in the clockwise direction for $Ra = 10^6, 10^8$, and $10^9$, whereas in the counterclockwise direction for $Ra = 5 \times 10^5$ and $10^7$ in our simulations. Therefore, for the consistency of our further analyses and discussions, we transform the horizontal velocity and temperature fields for $Ra = 5 \times 10^5$ and $10^7$ by reflecting them about the plane $x = L/2$, which changes the direction of LSC. As a result, the LSC rotates in the clockwise direction in all our simulations. 

We exhibit the instantaneous temperature field for $Ra = 10^6-10^9$ in figure~\ref{fig:flow_inst} along with the instantaneous velocity vectors. We observe that due to the decreasing Kolmogorov and Batchelor length scales with increasing $Re$ or $Ra$, the finest thermal structures in the flow, which are coarser for $Ra = 10^6$, become increasingly finer with increasing $Ra$. Also, the thickness of the thermal plumes, which are emitted mostly from the bottom left and top right parts of the plates decreases with increasing $Ra$. This is because the width of the thermal plumes is similar to the thickness of the thermal BLs~\citep{Zhou:PRL2007, Shishkina:JFM2008}, which decreases with increasing $Ra$. We also show the time-averaged flow structure for $Ra = 10^6-10^9$ in figure~\ref{fig:flow_mean}, which reveals that the mean LSC structure is octagonal (or circular) in our low-$Pr$ RBC. The corner flow structures become weaker and the LSC becomes increasingly squarish with increasing $Ra$ in our flow. The mean flow structure in our low-$Pr$ RBC is different from the mean flow structures in moderate- and high-$Pr$ 2D RBC for similar Rayleigh numbers, where the LSC is observed to be usually aligned along a diagonal of the cell~\citep{Sugiyama:PRL2010, Zhou:POF2011, Chandra:PRL2013, Zhang:PRE2017, Zhang:JFM2017}.

\begin{figure}
\centerline{
\includegraphics[width=\textwidth]{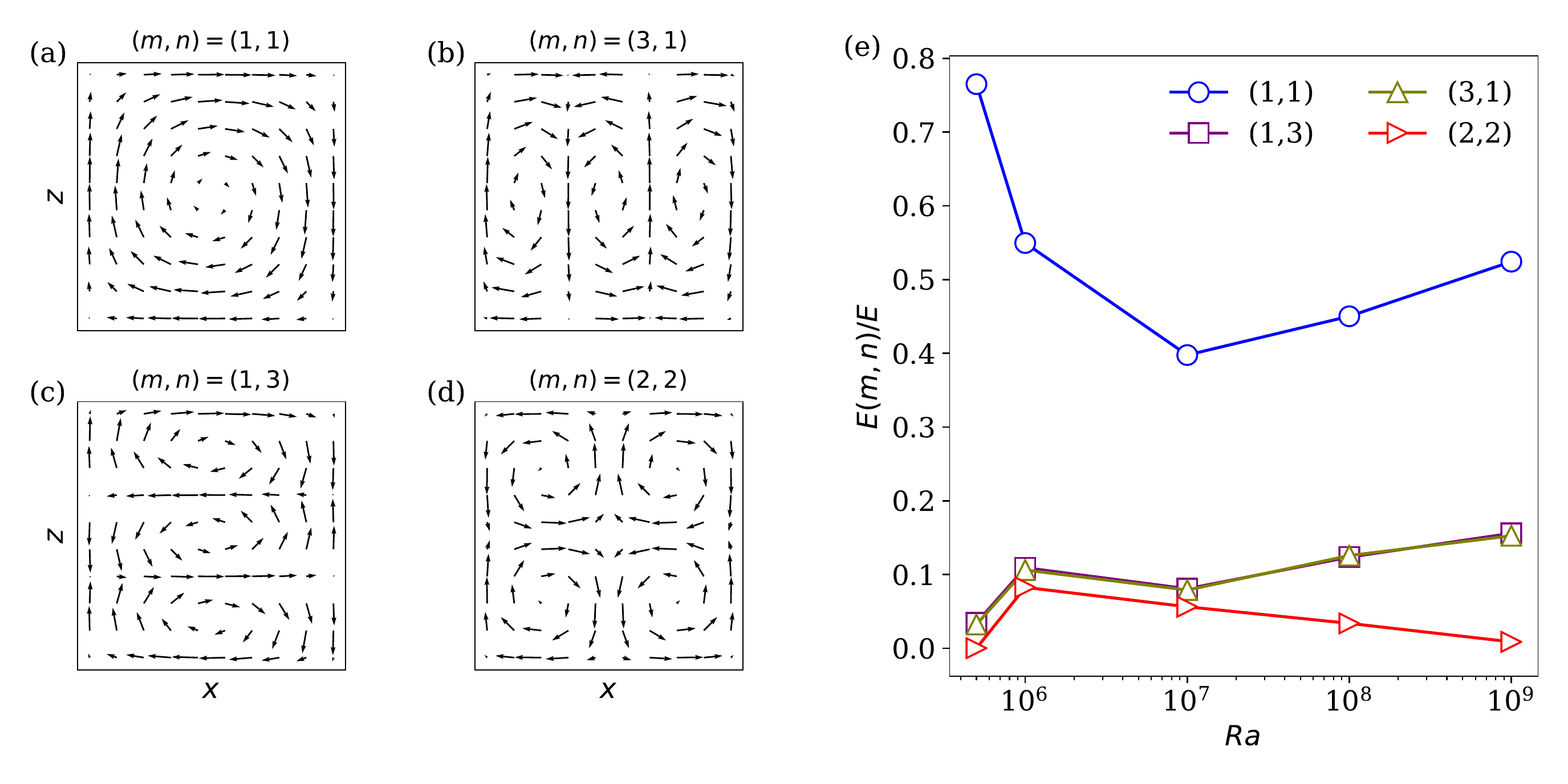}
}
\caption{Flow structures corresponding to the (a) $(1,1)$-mode, (b) $(3,1)$-mode, (c) $(1,3)$-mode, and (d) $(2,2)$-mode. (e) Fraction of the total kinetic energy contained in these flow modes as a function of $Ra$. In the turbulent regime, i.e., for $Ra \geq 10^7$, the strength of the LSC, which is represented by the $(1,1)$-mode, increases with increasing $Ra$. Moreover, strong corner flow structures, approximately represented by the (2,2)-mode, are present in the flow for $Ra = 10^6$.}
\label{fig:ke_modes}
\end{figure}

Figure~\ref{fig:flow_inst} also exhibits that, in addition to a large-scale structure, smaller structures are also present in the flow. The strengths of the flow structures of various sizes, also known as the flow modes, vary during the evolution of the flow due to the nonlinear interactions~\citep{Chandra:PRL2013}. See also supplementary movies exhibiting the temporal evolution of our low-$Pr$ flow for each $Ra$.
We estimate the strengths of various flow modes by computing their kinetic energy content by projecting the velocity field on a sine-cosine basis~\citep{Wagner:POF2013, Chen:JFM2019} as follows:
\begin{eqnarray}
u_x({\bm x},t) & = & \sum_{m,n} \hat{u}_x(m,n,t) [-2 \sin (m \upi x) \cos (n \upi z)], \\
u_z({\bm x},t) & = & \sum_{m,n} \hat{u}_z(m,n,t) [2 \cos (m \upi x) \sin (n \upi z)]. 
\end{eqnarray}
Here $m,n \in I$, i.e., they are positive integers. A flow mode with indices $(m,n)$ represents the flow structure with $m$ horizontally-stacked rolls and $n$ vertically-stacked rolls. Thus, the LSC is represented by the $(1,1)$-mode, and the smaller flow structures, such as the corner rolls, are represented by modes with higher indices~\citep{Chandra:PRL2013, Wagner:POF2013, Pandey:PRE2018, Chen:JFM2019}. We exhibit the flow structures corresponding to a few dominant modes in our flow in figure~\ref{fig:ke_modes}(a--d). 

The amplitude of the modes is computed as $\hat{u}_x(m,n) = \langle -2 u_x({\bm x})  \sin (m \upi x) \cos (n \upi z) \rangle_A$ and $\hat{u}_z(m,n) = \langle 2 u_z({\bm x})  \cos (m \upi x) \sin (n \upi z) \rangle_A$, and the kinetic energy contained in a mode $(m,n)$ is given as
\begin{equation}
E(m,n) = |\hat{u}_x(m,n)|^2 + |\hat{u}_z(m,n)|^2.
\end{equation}
We compute the time-averaged kinetic energy of various flow modes by considering $m, n = 1-10$, and plot the fraction of the total kinetic energy $E  = \langle u_x^2 + u_z^2 \rangle_{A,t}$ contained in a few strongest modes in our flow as a function of $Ra$ in figure~\ref{fig:ke_modes}(e).
Figure~\ref{fig:ke_modes}(e) reveals that for $Ra = 5 \times 10^5$, the flow is primarily dominated by one large-scale structure, which becomes weaker for $Ra = 10^6$ due to the growth of the smaller flow structures. This can be confirmed from the supplementary movies and from figure~\ref{fig:flow_inst}(a), which exhibit that strong corner flow structures (represented approximately by the (2,2)-mode) are present in the flow at $Ra = 10^6$. The strength of the LSC further decreases for $Ra = 10^7$ as even smaller structures are generated due to an increased Reynolds number of the flow. However, for $Ra \geq 10^7$, the strength of the LSC, i.e., of the (1,1)-mode grows and the total strength of the small-scale structures decays with increasing $Ra$~\citep{Sugiyama:PRL2010, Chandra:PRL2013}. This indicates that with increasing $Ra$ the LSC structure gets more and more squarish and increasingly fills the entire domain in our low-$Pr$ RBC~\citep{Lui:PRE1998, Niemela:EPL2003}.

\begin{figure}
\centerline{
\includegraphics[width=0.8\textwidth]{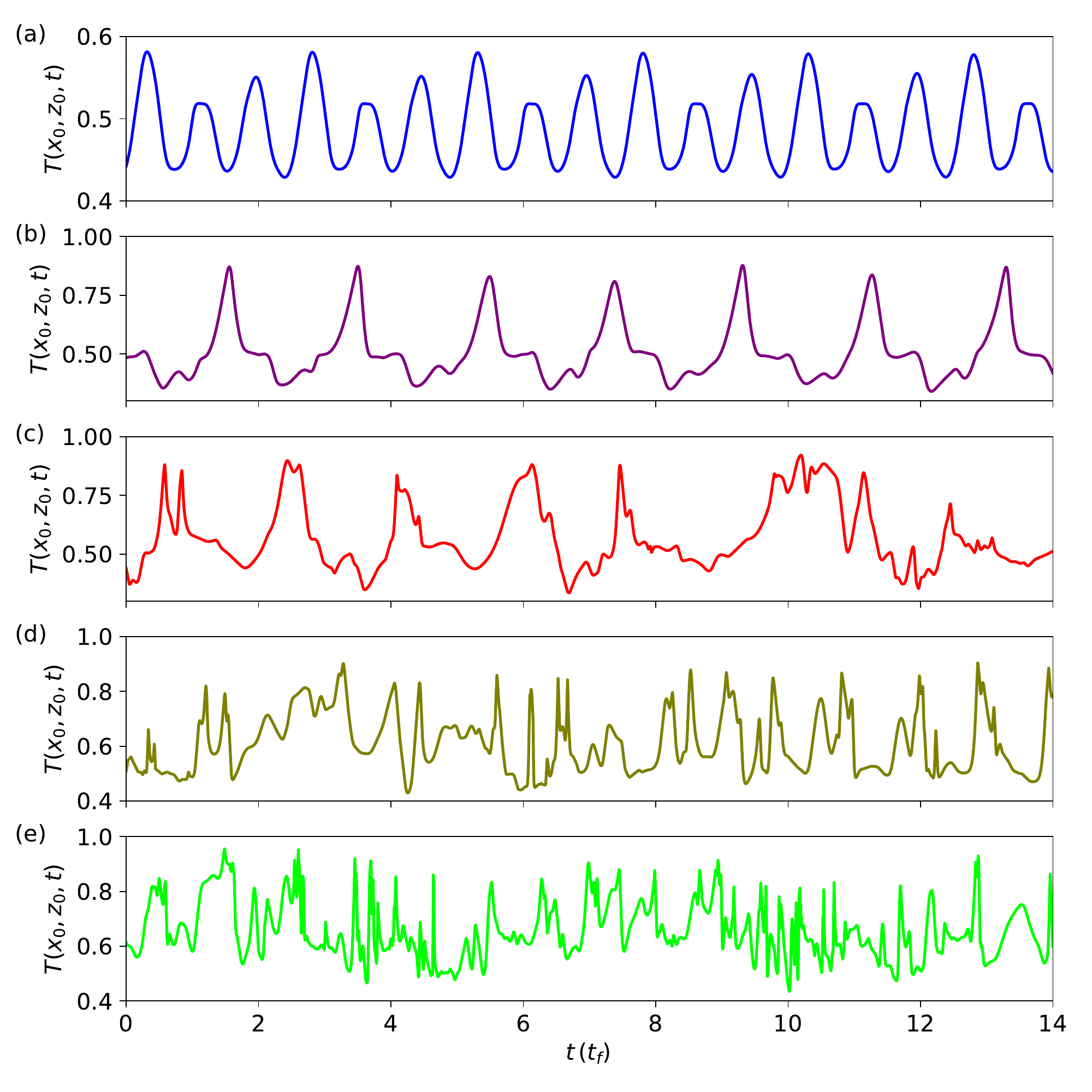}
}
\caption{A segment of the time traces of the temperature field at a fixed probe near the center of the bottom plate at the thermal BL height, i.e., at $x_0 = L/2, \,  z_0 \approx \delta_{\langle T \rangle}$ for (a) $Ra = 5 \times 10^5$, (b) $Ra = 10^6$, (c) $Ra = 10^7$, (d) $Ra = 10^8$, and (e) $Ra = 10^9$ (taken from the Run 5a). Panel (a) shows that the flow is periodic for $Ra = 5 \times 10^5$, whereas the flow is nearly periodic for $Ra = 10^6$, as indicated by the signal in panel (b). The temperature at the probe varies chaotically in panels (c--e), indicating that the flow is turbulent for $Ra \geq 10^7$. }
\label{fig:time_trace}
\end{figure}

Furthermore, we observe that the evolution of the flow for $Ra = 5 \times 10^5$ is unsteady and periodic time-dependent, and that for $Ra = 10^6$ is nearly periodic, i.e., some non-periodicity sets in the flow. To depict this, we look at the time trace of the temperature field at a fixed probe in the flow. Figure~\ref{fig:time_trace} shows a segment of the temperature traces at a probe located at the center of the bottom plate near the thermal BL height, i.e., at $x_0 = L/2, z_0 \approx \delta_{\langle T \rangle}$. Figure~\ref{fig:time_trace}(a) reveals that the temperature signal is periodic for $Ra = 5 \times 10^5$, whereas figure~\ref{fig:time_trace}(b) exhibits that some non-periodicity appears in the signal for $Ra = 10^6$. The temperature field at the probe for $Ra \geq 10^7$ vary chaotically, indicating that the flow for $Ra \geq 10^7$ is turbulent (see also supplementary movies).

\begin{figure}
\centerline{
\includegraphics[width=0.45\textwidth]{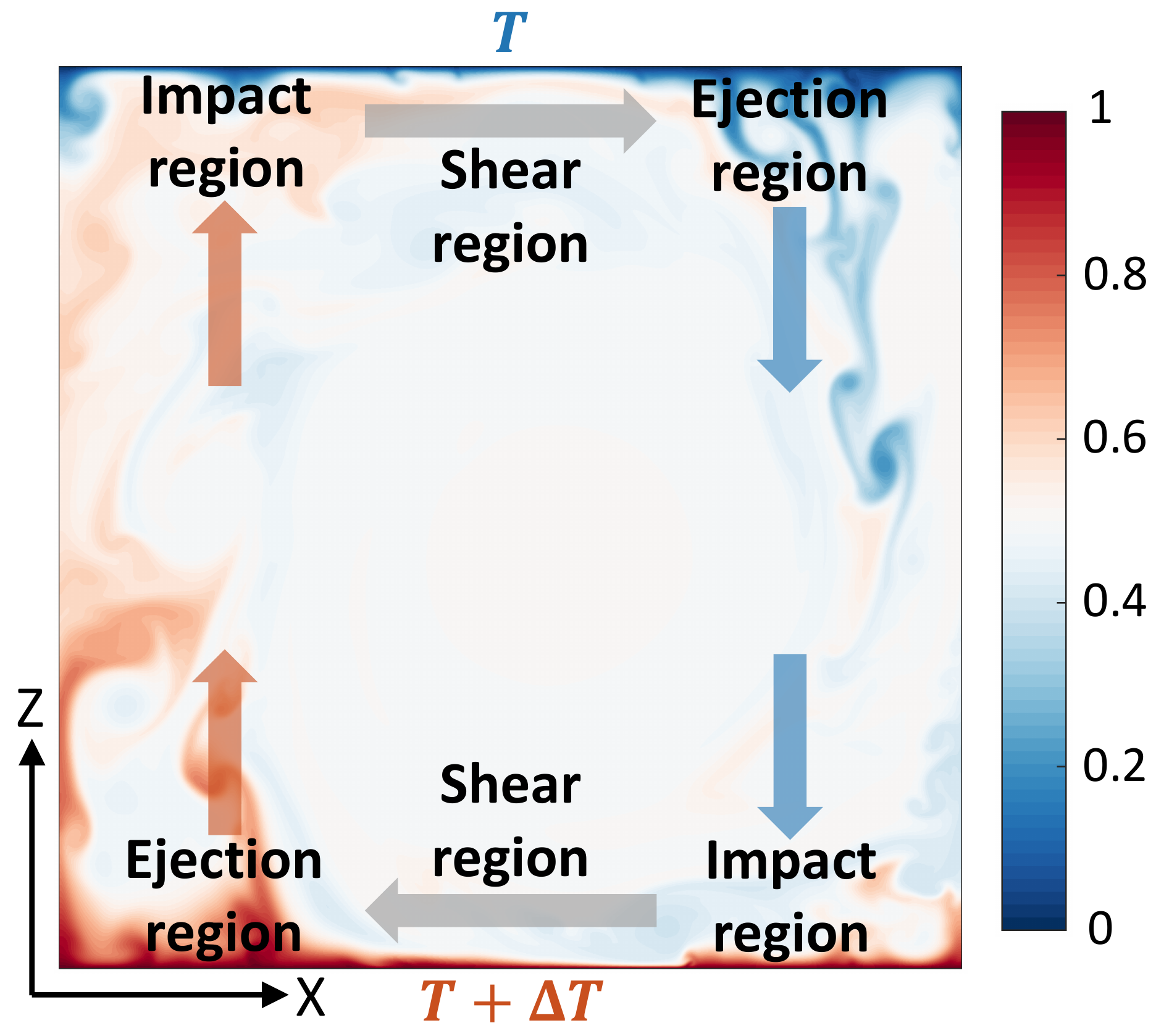}
}
\caption{A caricature exhibiting the plume-ejection, shear-dominated, and plume-impact regions at the top and bottom plates in our low-$Pr$ 2D RBC. An instantaneous snapshot of the temperature field for $Ra = 10^9$ is used to depict these regions.}
\label{fig:caricature}
\end{figure}

Unlike the flow in a pipe or channel, the RBC flow in a confined domain is not homogeneous in the horizontal directions. Instead, the horizontal plates can be divided into three distinct flow regions. For RBC in a domain of aspect ratio around unity, the hot plumes are generated primarily near one of the sidewalls and ascend towards the cold plate along that sidewall. Similarly, the cold plumes detaching from the top plate descend towards the hot plate along the opposite sidewall. These two regions are denoted respectively as the plume-ejection and plume-impact regions~\citep{Poel:PRL2015, Schumacher:PRF2016, Zhu:PRL2018}. In between these two regions, there exists a shear-dominated region, where the LSC is nearly parallel to the horizontal plates. We also depict these regions in our flow in a caricature in figure~\ref{fig:caricature}.
 Due to this specific flow morphology of RBC in a confined domain, the thermal and viscous BL profiles have often  been studied in the shear-dominated central region~\citep{Zhou:JFM2010, Poel:JFM2013, Schumacher:PRF2016, Wang:JFM2018}. In this study, however,  we compute the local temperature profiles in the aforementioned three regions and find that the BL profiles in the shear and impact regions are similar, and differ from the profiles in the ejection region. 


\section{Thermal boundary layer thicknesses}  

In this section, we discuss the scaling of the thermal BL width computed using the local as well as the horizontally-averaged temperature profiles. 

\subsection{Local boundary layer thicknesses}
We observe for all $Ra$  that the central region near the plates is dominated by shear generated due to LSC, and therefore, the flow properties at $x \approx L/2$ are similar to those in a shear flow.  The locations $x = L/4$ and $3L/4$ at the bottom plate in our flow approximately correspond to the plume-ejection and plume-impact regions, i.e., the hot plumes are ejected from the bottom plate mostly  at $x \approx L/4$ and the cold plumes impact at the bottom plate mostly at $x \approx 3L/4$ (see figure~\ref{fig:flow_inst}). The situation is reversed at the top plate, i.e., the physical location of the ejection region at the bottom corresponds to the impact region at the top, and vice versa. Also refer to figure~\ref{fig:caricature}, where we have summarized  these flow regions at the plates in a caricature. We have analyzed the properties of the temperature profiles measured at several locations in the flow. However, for clarity, we will mainly discuss the profiles measured at $x=L/4, L/2$, and $3L/4$ as these positions at the bottom plate typically correspond to the ejection, shear, and impact regions, respectively. We would like to point out that the large-scale structures in our flow fluctuate strongly in time (see supplementary movies), which causes the flow properties at the aforementioned locations to be occasionally influenced by the properties from the other regions. Therefore, we inevitably sample a mixed statistics due to the use of an immovable observational window in the flow.

\begin{figure}
\centerline{
\includegraphics[width=0.9\textwidth]{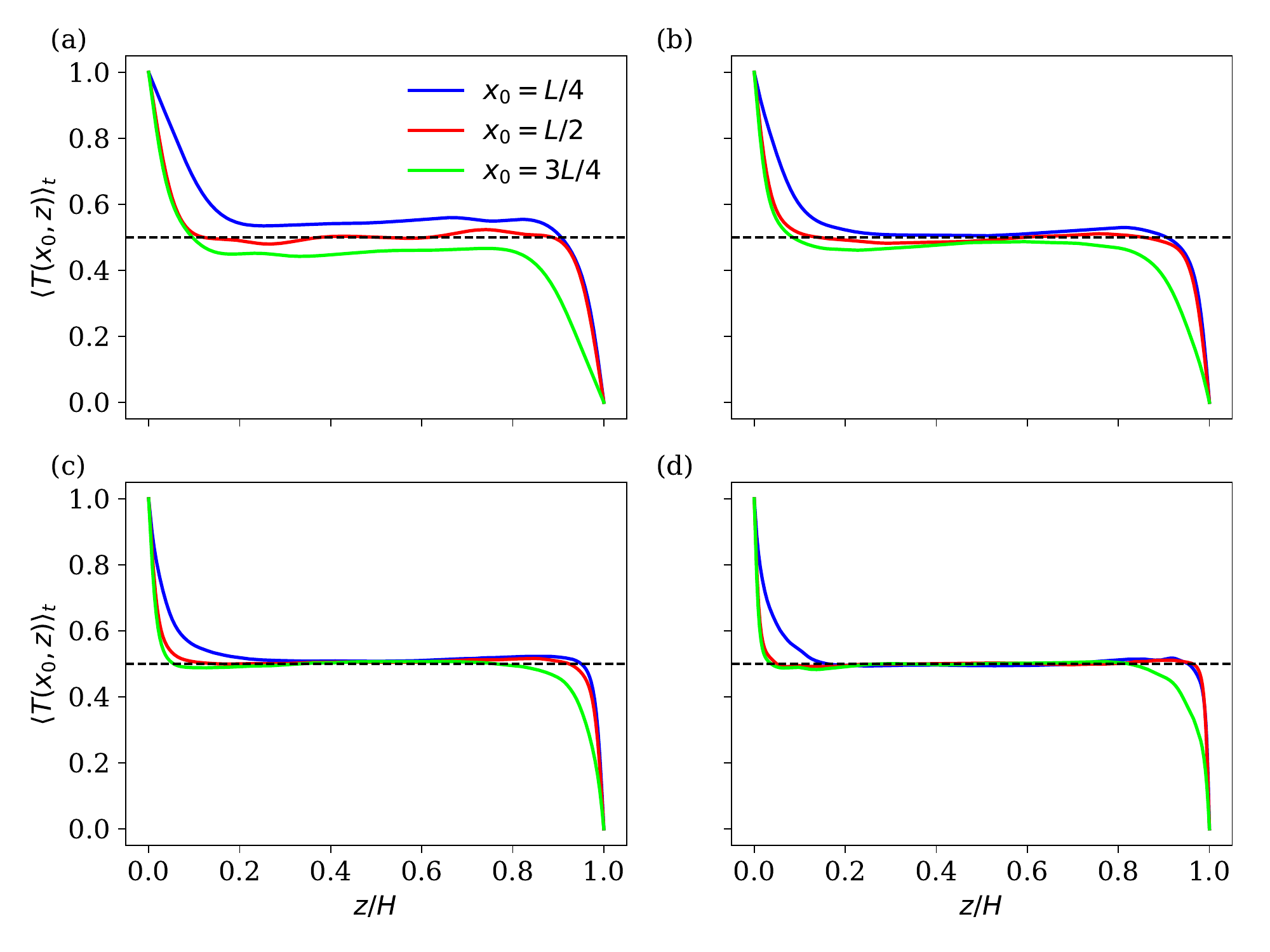}
}
\caption{Time-averaged temperature profiles measured at different horizontal positions for (a) $Ra = 10^6$, (b) $Ra = 10^7$, (c) $Ra = 10^8$, and (d) $Ra = 10^9$. For all $Ra$, profiles at $x=L/2$ are symmetric about the midplane, whereas those at $x \neq L/2$ do not exhibit symmetry due to different flow properties at the opposite plates at the particular physical location. In all the panels, blue curves near the bottom plate are similar to green curves near the top plate, and vice versa.}
\label{fig:T_z_local}
\end{figure}
We show the time-averaged temperature profiles at $x=L/4, L/2$, and $3L/4$ for $Ra = 10^6-10^9$ in figure~\ref{fig:T_z_local}. To reduce the scatter, we average the profiles in a tiny neighborhood around the aforementioned locations. Specifically, the profile at $x_0$ is averaged in the region corresponding to $x_0-0.02L \leq x \leq x_0+0.02L$, where $x_0 = L/4, L/2, 3L/4$. Figure~\ref{fig:T_z_local} shows that the profiles at $x = L/2$ are nearly symmetric about the midplane as $x \approx L/2$ corresponds to the shear-dominated region at both the top and bottom plates. The profiles at $x \neq L/2$, however, are not symmetric about the midplane. For instance, we can see in figure~\ref{fig:T_z_local}(d) for $Ra = 10^9$ that the profile near the bottom (top) plate at $x=L/4 \, (3L/4)$, which corresponds to the ejection region, approaches the bulk temperature slowly compared to the profiles in the other two regions. This is because the hot (cold) plumes that are ejected from the bottom (top) plate in the ejection region carry their thermal energy for a longer time and travel farther in the bulk region before losing their heat due to thermal diffusion. Moreover, the profile near the bottom (top) plate at $x = 3L/4 \, (L/4)$ corresponding to the impact region looks similar to that at $x=L/2$ in the shear region. Figure~\ref{fig:T_z_local} shows that the local profiles for all the Rayleigh numbers exhibit similar behaviour. Thus, to conclude, the temperature profiles in the impact and shear regions are similar, whereas that in the ejection region differs from them. Therefore, from now on, we will discuss the properties of the temperature profiles based on whether they belong to the impact, ejection, or shear regions, and not based on their physical location in the flow domain.

\begin{figure}
\centerline{
\includegraphics[width=0.9\textwidth]{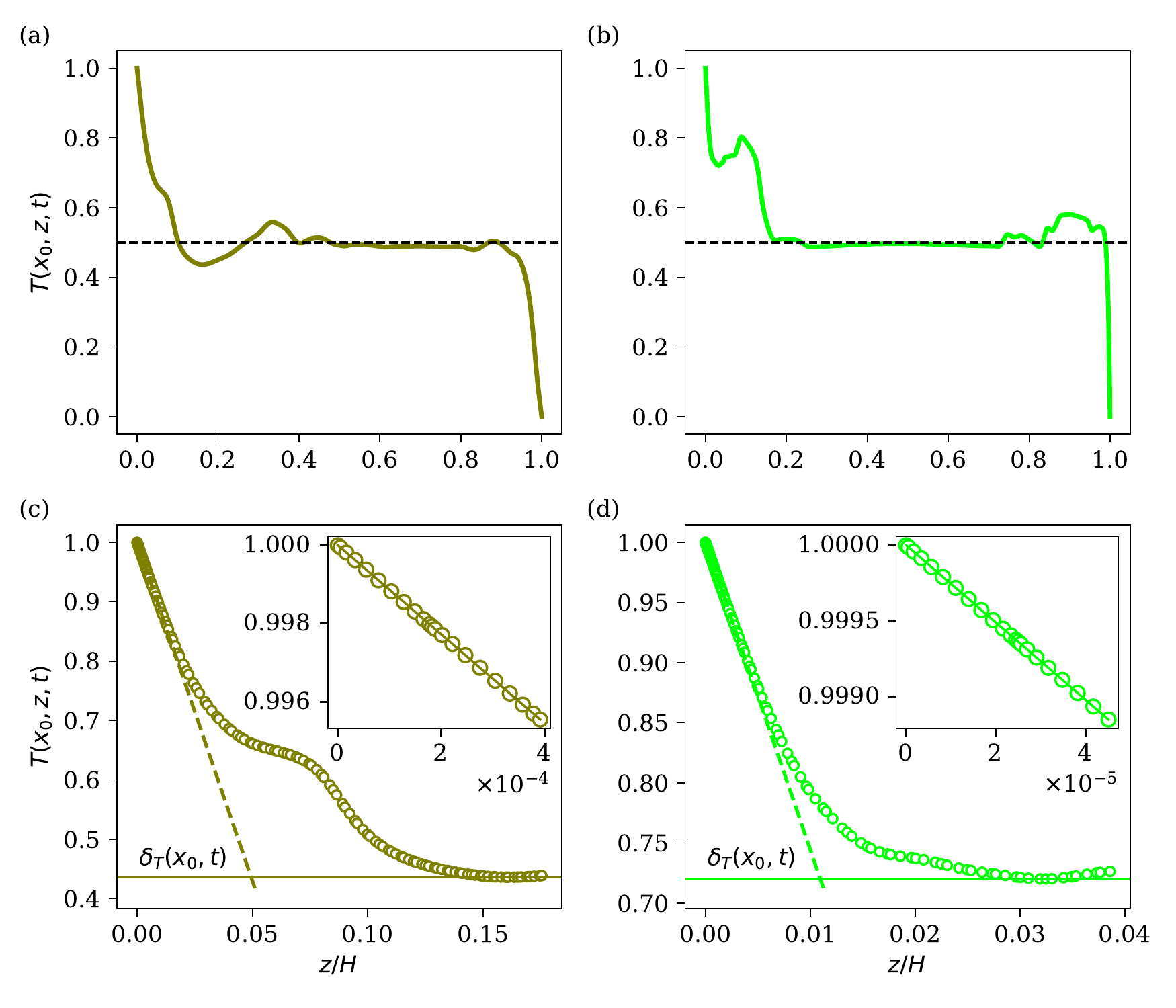}
}
\caption{Instantaneous local temperature profiles (a,b) and their magnification near the bottom plate (c,d) to demonstrate the determination of the instantaneous local thermal BL thickness using the slope method. (a,c) A profile measured in the shear-dominated region at $x_0 = L/2$ for $Ra = 10^8$ and (b,d) a profile measured in the plume-ejection region at $x_0 = L/4$ for $Ra = 10^9$. Position of the intersection of the horizontal line drawn at the first minimum with the slope at the plate yields $\delta_T(x_0,t)$. Panels (a,b) show that the temperature at the first minimum may differ from $T_\mathrm{bulk} = \Delta T/2$. Inset in panels (c,d) indicates the linear variation of temperature in the vicinity of the bottom plate.}
\label{fig:compute_delta}
\end{figure}

We compute the local thermal BL width $\delta_T(x)$ using the slope method~\citep{Zhou:POF2011, Scheel:JFM2012, Wagner:JFM2012}, where $\delta_T(x)$ is estimated as the distance from the plate where the slope of the temperature profile drawn at the plate meets the horizontal line passing through the first minimum of the profile. In figure~\ref{fig:compute_delta}, we demonstrate this method of determining the BL thickness from the instantaneous temperature profiles measured at two different positions at the plate. Figure~\ref{fig:compute_delta} exhibits that the profiles do not always approach the bulk temperature monotonically and the first minimum may differ from the bulk temperature. We also compute the instantaneous BL thicknesses 
using the local vertical temperature gradient at the plate as $\delta_T(x) = 0.5/|\partial T(x)/\partial z|_{z=0}$, which is equivalent to computing the BL thickness using the slope method with the horizontal line drawn at the mean temperature $\Delta T/2$. We find, however, that the mean (and most of the time instantaneous) BL thicknesses determined using this method are nearly the same as those obtained using the first minimum method. We, however, prefer the first minimum method to include the instantaneous variation of the profiles as the temperature in the bulk region may instantaneously differ from $\Delta T/2$.

\begin{figure}
\centerline{
\includegraphics[width=\textwidth]{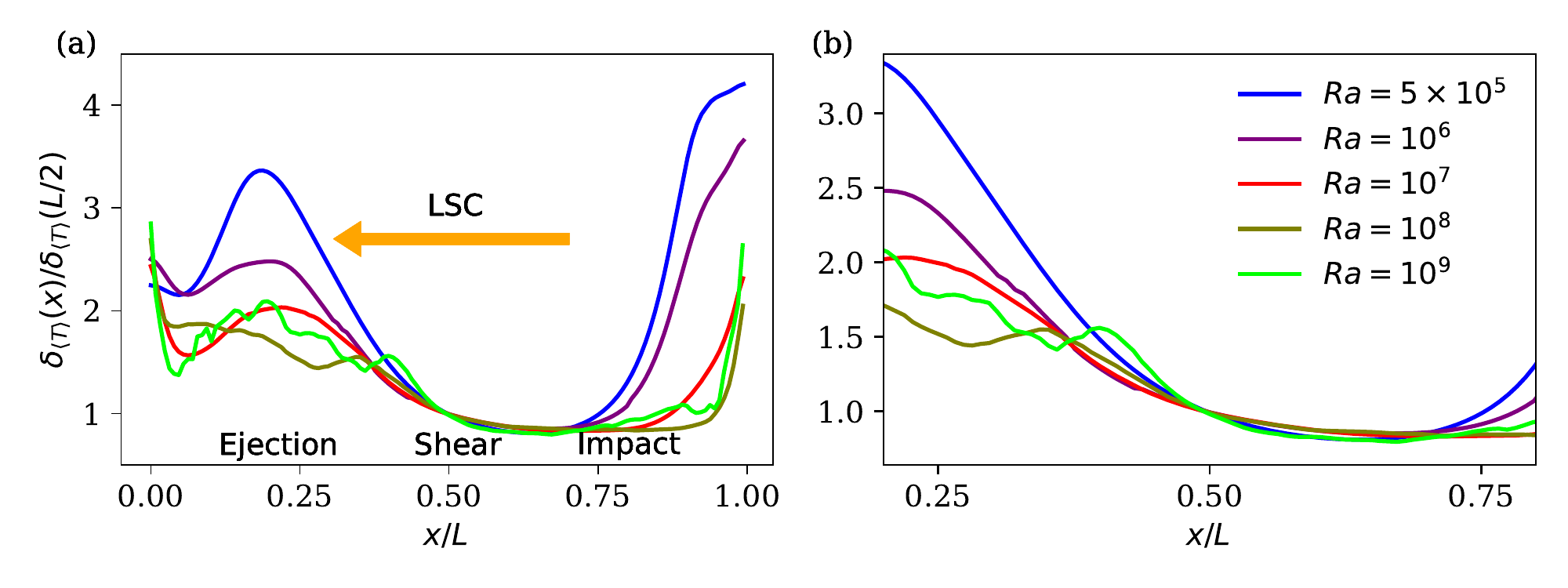}
}
\caption{(a) Variation of the normalized local BL thicknesses along the horizontal plate computed using the time-averaged profiles. Except very close to the sidewalls, the local thicknesses $\delta_{\langle T \rangle}(x)$ are larger in the ejection region and decrease as the impact region is approached, i.e.,  the thickness grows in the direction of LSC. (b) Magnification of panel (a) exhibiting the BL structure in the central region away from the sidewalls shows that the BL structure is nearly independent of $Ra$ in the impact and shear regions.}
\label{fig:blw_x}
\end{figure}

To explore the structure of the thermal BL, we compute the local BL thicknesses  $\delta_{\langle T \rangle}(x)$ at both the plates using the time-averaged temperature profiles. Figure~\ref{fig:blw_x} shows the local BL width averaged over the top and bottom plates and normalized with the BL width at $x=L/2$. Note that due to the interchange of the ejection and impact regions, local BL thickness at a position $x$ in figure~\ref{fig:blw_x} is the average of $\delta_{\langle T \rangle}(x)$ at the bottom and $\delta_{\langle T \rangle}(L-x)$ at the top plate. Figure~\ref{fig:blw_x} reveals that  the BL structure is similar for all the explored $Ra$ in our low-$Pr$ RBC.  We observe that $\delta_{\langle T \rangle}(x)$ is the largest near the sidewalls and decreases as one moves towards the central region. Moreover, the local thicknesses are larger in the ejection region than those in the other two regions, which are indicated in figure~\ref{fig:blw_x}(a). We also find that $\delta_{\langle T \rangle}(x)$ in the impact region are a bit smaller than those in the shear region. To show the BL structure away from the sidewalls, we plot $\delta_{\langle T \rangle}(x)$ for $0.2L \leq x \leq 0.8L$ in figure~\ref{fig:blw_x}(b), which reveals that the BL structure is nearly independent of $Ra$ in the impact and shear regions. In the ejection region, however, the variation of the normalized BL thickness depends on $Ra$. We quantify the relative variation of the BL thicknesses exhibited in figure~\ref{fig:blw_x}(b) as 
\begin{equation}
\mc{R} = \frac{[\delta_{\langle T \rangle}(x)]_\mathrm{max}-[\delta_{\langle T \rangle}(x)]_\mathrm{min}}{[\delta_{\langle T \rangle}(x)]_\mathrm{min}}, 
\end{equation}
where $[\delta_{\langle T \rangle}(x)]_\mathrm{max}$ and $[\delta_{\langle T \rangle}(x)]_\mathrm{min}$ are respectively the maximum and minimum BL thicknesses over the central region, i.e., in the region $0.2L \leq x \leq 0.8L$ at the plate. We find $\mc{R} = 3.1, 1.9, 1.5, 1.0, 1.6$ respectively for $Ra = 5 \times 10^5, 10^6, 10^7, 10^8,10^9$, which indicates that $\mc{R}$ generally decreases with increasing $Ra$ in our flow. Our data at $Ra = 10^9$ does not follow the decreasing trend very well, which might be due to a limited statistics available for this $Ra$. A few more simulations at intermediate Rayleigh numbers would probably yield a better picture of the BL structure in the ejection region in our low-$Pr$ flow.

According to the Prandtl-Blasius BL theory the BL thickness grows as $\sqrt{x}$ in the downstream direction for a laminar BL, whereas grows linearly with $x$ for a turbulent BL~\citep{Schlichting:book2004}. In our 2D RBC, the hotter or colder fluid impinges on the plate in the impact region and then moves along the plate towards the ejection region. We thus also observe a growth of the BL thickness in the downstream direction, i.e., in the direction of the LSC on the plate. However, $\delta_{\langle T \rangle}(x)$ in our low-$Pr$ RBC does not follow either of the aforementioned scalings. Possible reasons for a different BL structure in our flow might be the assumptions in the PB BL theory, which are not totally satisfied in our 2D convection flow. For example, the BLs in our flow are not entirely laminar and the turbulent fluctuations within the BL region become stronger with increasing $Ra$ (see figure~\ref{fig:kappa_t}(b)). We moreover observe that the BL structure is not symmetric about the center of the plate, which is qualitatively similar to those observed in high-$Pr$ RBC in 2D~\citep{Werne:PRE1993, Zhou:POF2011}. The BL structure in our low-$Pr$ convection is also qualitatively similar to the observations of \citet{Wagner:JFM2012} in a cylindrical cell of aspect ratio one that the BL width increases in the direction of LSC as well as to the observations of \citet{Scheel:JFM2014} that the local BL widths are larger in the plume-ejection region. \citet{Wagner:JFM2012}, however, observed a nearly linear growth of the BL thickness along the direction of LSC. Our results are also different from those of~\citet{Lui:PRE1998}, who studied the thermal BL structure in a cylindrical cell of aspect ratio one filled with water and observed that, in the plane of LSC, the BL width is minimum at the center of the plate and increases symmetrically towards the sidewalls. The aforementioned differences between the BL structure in our 2D RBC with those observed in 3D RBC, therefore, indicate that the quasi-2D nature along with the other characteristics of LSC in 3D RBC also affects the BL structure.

\begin{figure}
\centerline{
\includegraphics[width=\textwidth]{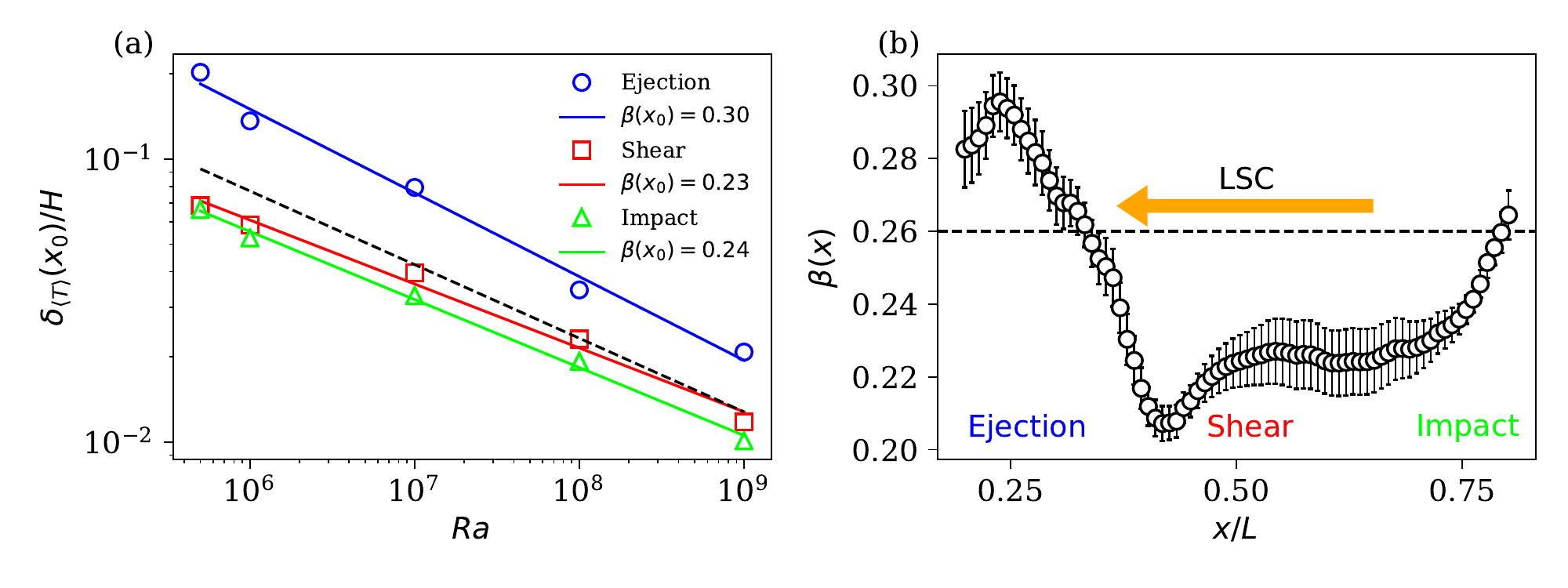}
}
\caption{(a) Local thermal BL thicknesses averaged over both the plates vary as $Ra^{-\beta(x)}$ in the ejection, shear, and impact regions, with $\beta(x_0)$ being larger in the ejection region. The black dashed line indicates the scaling of the BL thickness $\delta_{\langle T \rangle}$ computed using the horizontally- and temporally-averaged profiles (shown in figure~\ref{fig:mean_T_z}(b)). If these scalings hold for the larger Rayleigh numbers, the blue line will intersect the red or green line at $Ra^* \approx 8 \times 10^{12}$, beyond which the thermal BL might become uniform at the plate. (b) Variation of the local scaling exponent $\beta(x)$ {along the plate} obtained from the BL thicknesses averaged over the top and bottom plates. The dashed horizontal line indicates the scaling exponent of the mean BL thickness. The direction of LSC and the ejection, shear, and impact regions are also indicated.}
\label{fig:delta_T_local}
\end{figure}

We compute the local thermal BL thicknesses $\delta_{\langle T \rangle}(x_0)$ at $x_0 = L/4, L/2,$ and $3L/4$ using the time-averaged profiles exhibited in figure~\ref{fig:T_z_local} and find that $\delta_{\langle T \rangle}(x_0)$ decreases with increasing $Ra$ as $Ra^{-\beta(x_0)}$, with the exponent $\beta(x_0)$ depending on the position at the plate. In figure~\ref{fig:delta_T_local}(a), we plot $\delta_{\langle T \rangle}(x_0)$ averaged over the corresponding regions at the top and bottom plates as a function of $Ra$, which reveals that the BL thicknesses in the impact and shear regions are similar for all $Ra$. This is consistent with the fact that the profiles in the shear and impact regions exhibited in figure~\ref{fig:T_z_local} are also similar. The best fit yields that $\delta_{\langle T \rangle}(x_0)/H$ varies as $(1.4 \pm 0.2)Ra^{-0.23 \pm 0.01}$ in the shear region and as $(1.5 \pm 0.1)Ra^{-0.24 \pm 0.01}$ in the impact region. The local BL thicknesses in the ejection region, however, are larger than those in the other two regions for all $Ra$,  and scale as $(8.9 \pm 1.1)Ra^{-0.30 \pm 0.02}$.

\begin{figure}
\centerline{
\includegraphics[width=\textwidth]{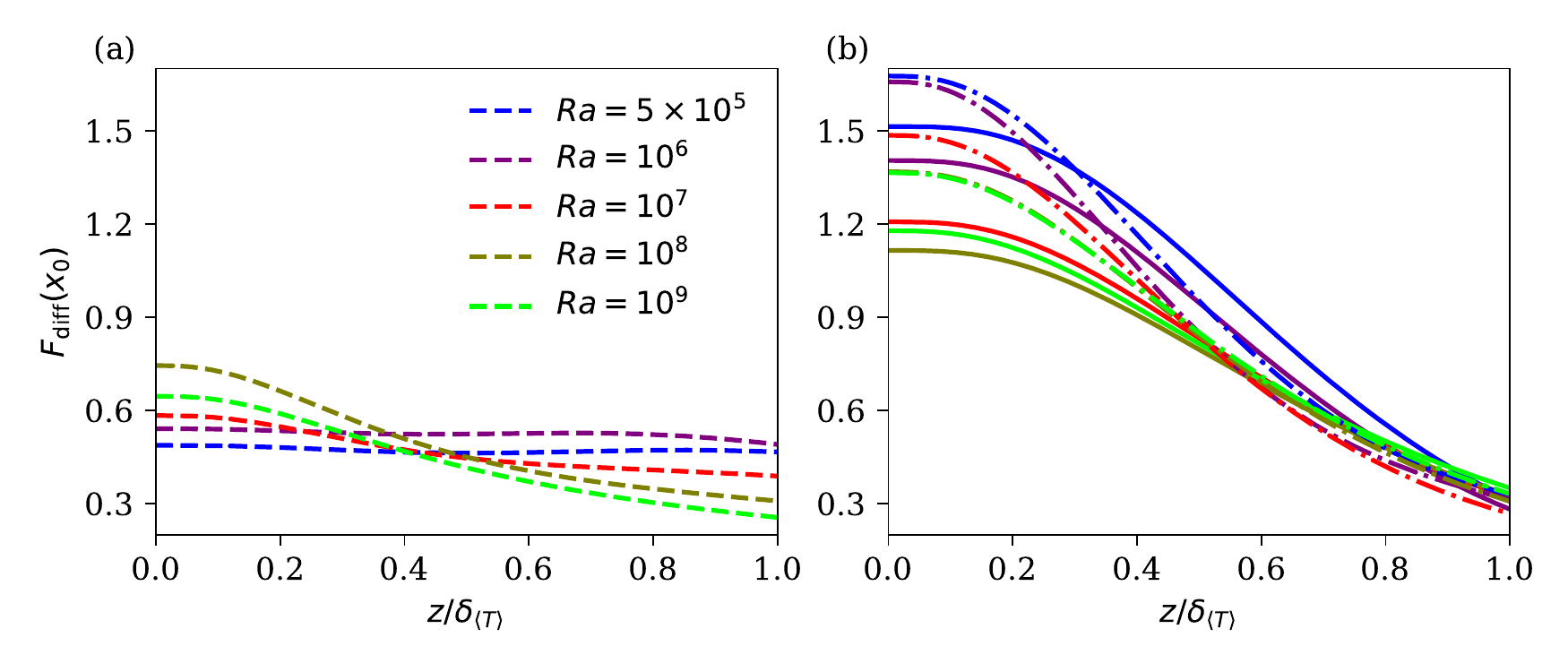}
}
\caption{Vertical profiles of the fraction $F_\mathrm{diff}(x_0)$ of the total heat flux carried by the thermal diffusion in the (a) ejection, (b) shear (solid curves), and impact (dash-dot curves) regions within the BL. $F_\mathrm{diff}(x_0)$ at the plate is smaller in the ejection region than those in the shear and impact regions, and with increasing $Ra$, $F_\mathrm{diff}(x_0)$ is generally increasing in the ejection region and decreasing in the other two regions.}
\label{fig:flux_fraction}
\end{figure}

We further explore the variation of the local exponent $\beta(x)$ at the plate by computing them from the local BL thicknesses. Figure~\ref{fig:delta_T_local}(b) exhibits the local exponents $\beta(x)$ computed from the averaged BL thicknesses at both the plates in the central region far from the sidewalls. We observe that the $\beta(x)$ are larger in the ejection region than those in the shear and impact regions, where they are nearly the same. Note that the thermal BL thickness is related to the diffusive heat flux at the plate. To understand the position-dependent variation of the properties of the thermal BL, we compute the diffusive fraction of the total heat flux as $F_\mathrm{diff}(x_0) = (H \partial \langle T \rangle_t(x_0)/\partial z)/(Nu \Delta T)$ in the ejection, shear, and impact regions. $F_\mathrm{diff}(x_0)$ as a function of $z/\delta_{\langle T \rangle}$ in the ejection region is exhibited in figure~\ref{fig:flux_fraction}(a), whereas those in the shear (solid curves) and the impact (dash-dot curves) regions are depicted in figure~\ref{fig:flux_fraction}(b). We observe for all $Ra$ that $F_\mathrm{diff}(x_0)$ at the plate and in its vicinity is smaller in the ejection region than those in the shear region, which in turn, are a bit smaller compared to those in the impact region. This is because of a larger temperature gradient in the impact region due to impinging cold (hot) plumes at the bottom (top) plate. As the BL width is inversely proportional to the vertical temperature gradient (or the diffusive flux) at the plate, this implies that the local BL thickness in the ejection region is larger compared to those in the shear and impact regions, which is consistent with the BL structure from figure~\ref{fig:blw_x}. 

Furthermore, we observe that $F_\mathrm{diff}(x_0)$ in the vicinity of the plate is  generally  increasing in the ejection region, whereas decreasing in the shear and impact regions with increasing $Ra$. This means that, with increasing $Ra$, the BL width in the ejection region decreases faster, whereas those in the other two regions decrease slower compared to the mean BL thickness, which agrees with the observations of figure~\ref{fig:delta_T_local}. Therefore, for moderately large Rayleigh numbers in our low-$Pr$ flow, the difference between the diffusive contributions from various regions is decreasing or, in other words, the variation of the diffusive heat flux over the plate becomes weaker with increasing $Ra$. The reason for this weaker variation is the increasing strength of LSC with increasing $Ra$ in our flow~\citep{Lui:PRE1998} as we have observed that the (1,1)-mode representing the LSC structure becomes stronger as $Ra$ increases (see figure~\ref{fig:ke_modes}(e)). A stronger LSC causes the cold (hot) plumes impinging on the bottom (top) plate in the impact region to move a larger distance along the plate before their heat is lost due to thermal diffusion. This causes an increasingly uniform temperature gradient along the plate with increasing $Ra$. If the observed trend in our flow continues to hold for larger $Ra$, the local heat flux in the ejection region might take over those in the other two regions for large enough $Ra$. This picture would be consistent with the findings of \citet{Zhu:PRL2018} in 2D RBC for $Pr = 1, Ra > 10^{11}$ that the local heat flux at the plate is larger in the ejection region. Figure~\ref{fig:flux_fraction} additionally shows that the variation of $F_\mathrm{diff}(x_0)$ for $z \geq 0.4 \delta_{\langle T \rangle}$ becomes nearly independent of $Ra$ in the shear and impact regions, whereas $F_\mathrm{diff}(x_0)$ generally decreases with increasing $Ra$ in the ejection region. This, in turn, suggests that away from the plate but within the BL region, the turbulent fraction of the heat flux in the ejection region increases with increasing $Ra$. Combining the above scenarios, our results suggest that in the whole BL the local heat flux in the ejection region becomes stronger with increasing $Ra$ in our low-$Pr$ convection.

Thus, the difference between the local thicknesses in the ejection and the other two regions decreases with increasing $Ra$, and, if the present trends hold also for the larger Rayleigh numbers, the local thicknesses for sufficiently large $Ra$ might become independent of the horizontal position. We can estimate this asymptotic $Ra$ by finding the intersection point of the blue line with either the red or the green line in figure~\ref{fig:delta_T_local}. We find that these lines will intersect at $Ra^* \approx 8 \times 10^{12}$, and therefore, the thermal BL structure in our low-$Pr$ convection might become uniform for $Ra \geq Ra^*$. By considering the error bars into account, $Ra^*$ may vary between $5 \times 10^{9}$ to $2 \times 10^{19}$, and thus the predicted range of $Ra^*$ is very wide. Note that the estimated $Ra^*$ is very large and would probably correspond to the {\it ultimate regime of convection} for this Prandtl number~\citep{Scheel:PRF2017}. \cite{Scheel:PRF2017} used various methods to predict the onset of the ultimate regime in a cylindrical cell of aspect ratio one and found interestingly that the onset for $Pr = 0.021$ may occur between $Ra = 5 \times 10^9$ and $10^{18}$. \citet{Lui:PRE1998} and \citet{Wang:EPJB2003} studied the thermal BL structure in water for high Rayleigh numbers and  also observed that in the direction of LSC the variation of the local BL thickness with the horizontal position becomes weaker with increasing $Ra$. 

In the next section, we discuss the scaling of the mean BL thickness at the plates.

\subsection{Mean boundary layer thickness}

\begin{figure}
\centerline{
\includegraphics[width=\textwidth]{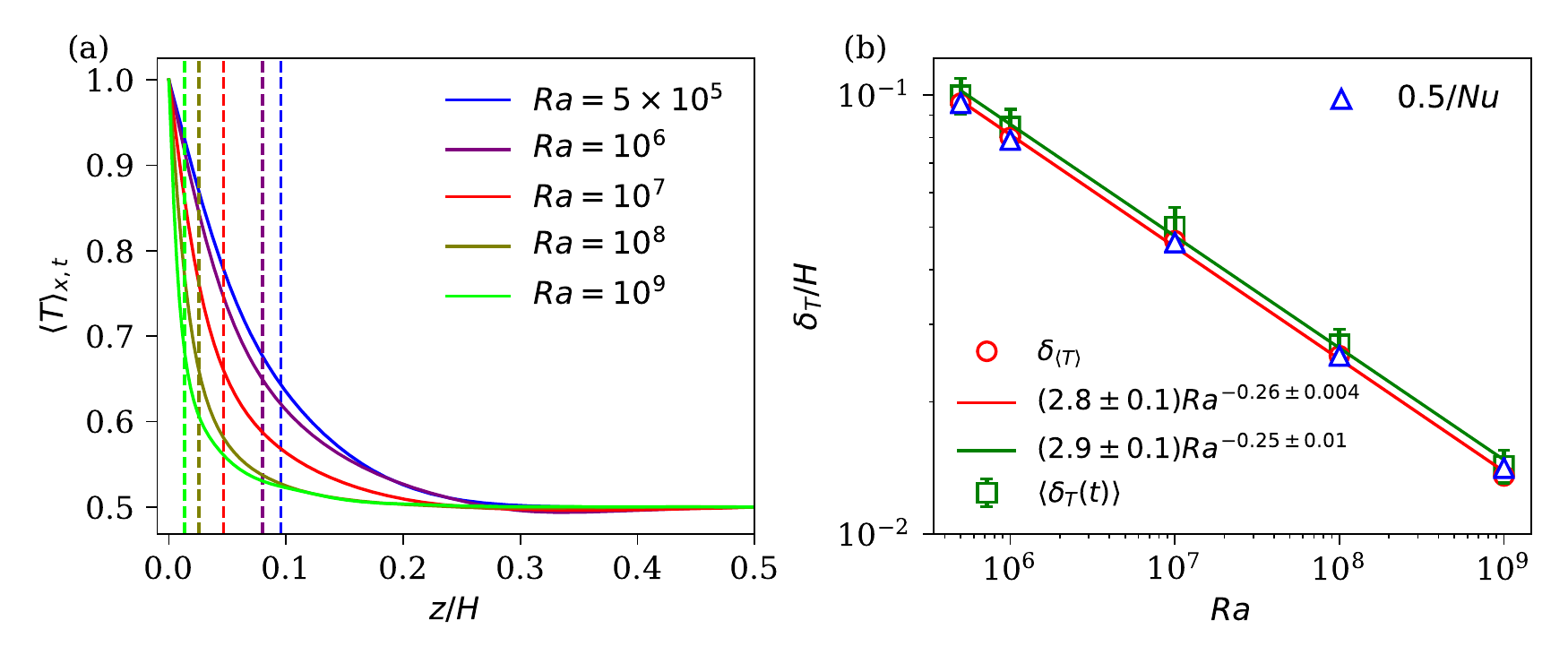}
}
\caption{(a) Horizontally- and temporally-averaged temperature profiles for all the Rayleigh numbers exhibit that the bulk fluid is mixed well and $T_\mathrm{bulk} \approx \Delta T/2 = 0.5$. (b) Mean thickness of the top and bottom thermal BLs decreases as $Ra^{-0.26}$. We show the thickness $\delta_{\langle T \rangle}$ computed from the time-averaged profiles as well as the time-averaged thickness $\langle \delta_T(t) \rangle$ computed from the instantaneous profiles with the errorbars in $\langle \delta_T(t) \rangle$ indicating the standard deviation of $\delta_T(t)$. Mean thermal BL thicknesses computed from the relation $\delta_{\langle T \rangle} = 0.5H/Nu$ are also indicated as blue triangles. Dashed vertical lines in panel (a) indicate $\delta_{\langle T \rangle}$ for all $Ra$ with the corresponding colors. }
\label{fig:mean_T_z}
\end{figure}

In RBC, the horizontally-averaged temperature varies primarily only in the thin thermal BLs and remains almost a constant in the bulk region. We compute the horizontally- and temporally-averaged temperature $\langle T \rangle_{x,t}$ for all $Ra$ and plot them in figure~\ref{fig:mean_T_z}(a), which exhibits that the mean temperature in the bulk is indeed approximately a constant. Figure~\ref{fig:mean_T_z}(a) shows the profiles averaged over the bottom and top halves of the domain, as they are symmetric about the midplane ($z=0.5H$) due to the Oberbeck-Boussinesq convection in the present case. We observe that the profiles approach the bulk temperature increasingly faster as $Ra$ increases, thus indicating that the diffusive region, where the vertical temperature gradient is significant, shrinks with increasing $Ra$. 

We compute the mean thermal BL width $\delta_{\langle T \rangle}$ from $\langle T \rangle_{x,t}$ using the slope method, and plot the average thickness of the top and bottom BLs as a function of $Ra$ in figure~\ref{fig:mean_T_z}(b), which shows that the average thickness decreases with increasing $Ra$  as $\delta_{\langle T \rangle}/H = (2.8 \pm 0.1)Ra^{-0.26 \pm 0.004}$. As aforementioned, the mean thermal BL thickness is related to $Nu$ as $\delta_{\langle T \rangle}/H = 0.5/Nu$. This is because the diffusive  contribution to the total heat transport decreases with increasing $Nu$, and thus, the region where diffusive processes are dominant shrinks as $Nu$ increases. Therefore, we also show the BL thickness computed as $0.5/Nu$ in figure~\ref{fig:mean_T_z}(b), and observe excellent agreement with those computed using the slope method. We also compute the thermal BL widths $\delta_T(t)$ from the instantaneous horizontally-averaged temperature profiles and average them to obtain the mean width $\langle \delta_T(t) \rangle$, which are also plotted in figure~\ref{fig:mean_T_z}(b). We observe that $\langle \delta_T(t) \rangle$ is slightly larger than $\delta_{\langle T \rangle}$ for all $Ra$, but its variation with $Ra$ is nearly the same. The best fit yields $\langle \delta_T(t) \rangle/H = (2.9 \pm 0.1)Ra^{-0.25 \pm 0.01}$, which agrees very well with $\delta_{\langle T \rangle}$ computed from the time-averaged profiles. 

\begin{figure}
\centerline{
\includegraphics[width=\textwidth]{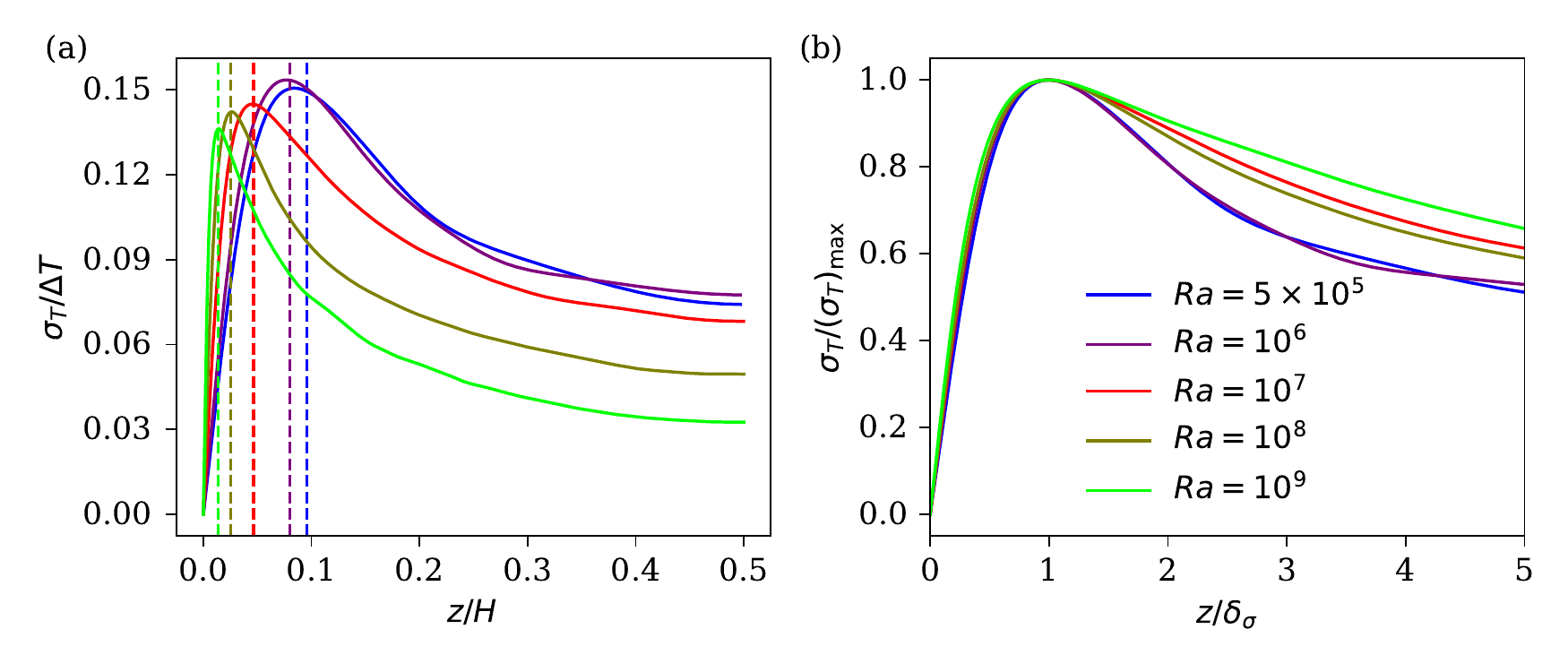}
}
\caption{(a) Vertical profiles of the rms temperature fluctuations averaged over the top and bottom halves of the domain for all $Ra$ attain their maximum value near the edge of the thermal BL and then decay monotonically in the bulk region towards the center. Dashed vertical lines with the corresponding colors indicate the mean BL thicknesses $\delta_{\langle T \rangle}$. (b) Profiles normalized with their maximum value as a function of the normalized vertical distance collapse reasonably well within the thermal BL region. }
\label{fig:sigma_T}
\end{figure}

The BLs in our low-$Pr$ RBC are not completely laminar and exhibit fluctuations for all the Rayleigh numbers. Therefore, we compute the rms temperature fluctuations from the mean temperature profile $\langle T \rangle_{x,t}$ as
\begin{equation}
\sigma_T(z) = \sqrt{ \langle (T - \langle T \rangle_{x,t})^2 \rangle_{x,t} }
\end{equation}
and plot $\sigma_T(z)$ averaged over the top and bottom halves of the domain in figure~\ref{fig:sigma_T}(a). We observe that the rms fluctuations increase with increasing distance from the plate and attain their maximum value near the edge of the thermal BL~\citep{Deardorff:JFM1967, Wang:EPJB2003, Zhou:JFM2013}. This is due to generation and perpetual emission of the thermal plumes inside the thermal BL. Note that the thickness of the plumes is similar to the thickness of the thermal BL and their temperature is higher compared to the ambient fluid within the bottom thermal BL~\citep{Zhou:PRL2007, Shishkina:JFM2008}. Therefore, the decrease of the horizontally-averaged temperature with the increasing distance from the bottom plate is primarily due to a decreasing temperature of the ambient fluid, as the plumes almost retain their heat or temperature within the BL region. Thus, the disparity between the temperatures of the ambient fluid and the plumes increases with increasing distance from the bottom plate, which causes the increase of $\sigma_T(z)$ within the thermal BL. A similar difference between the temperatures of the cold plumes and the ambient fluid in the top thermal BL causes the increase of $\sigma_T(z)$ near the top plate. After attaining the maximum, the rms fluctuations decline monotonically as the central region of the flow is approached. This is because the plumes are not able to retain their temperature due to an increased turbulent mixing outside the BL region.

Therefore, it is clear from figure~\ref{fig:sigma_T}(a) that the position of the maximum of $\sigma_T$ also yields a measure of the BL thickness~\citep{Wang:EPJB2003, Zhou:JFM2013}. We observe that $\delta_\sigma$, the position corresponding to the maximum of $\sigma_T$, as well as the maximum amplitude of the fluctuations, decrease with increasing $Ra$. We therefore show the normalized temperature variance profiles $\sigma_T/(\sigma_T)_\mathrm{max}$ as a function of the normalized distance $z/\delta_\sigma$ in figure~\ref{fig:sigma_T}(b), and find that the normalized profiles collapse reasonably well within the BL region, i.e., up to $z \approx \delta_\sigma$~\citep{Zhou:JFM2013}. However, the profiles do not seem to collapse outside the  BL region. This indicates that $\delta_\sigma$ is indeed a characteristic length scale within the thermal BL region.
\begin{figure}
\centerline{
\includegraphics[width=\textwidth]{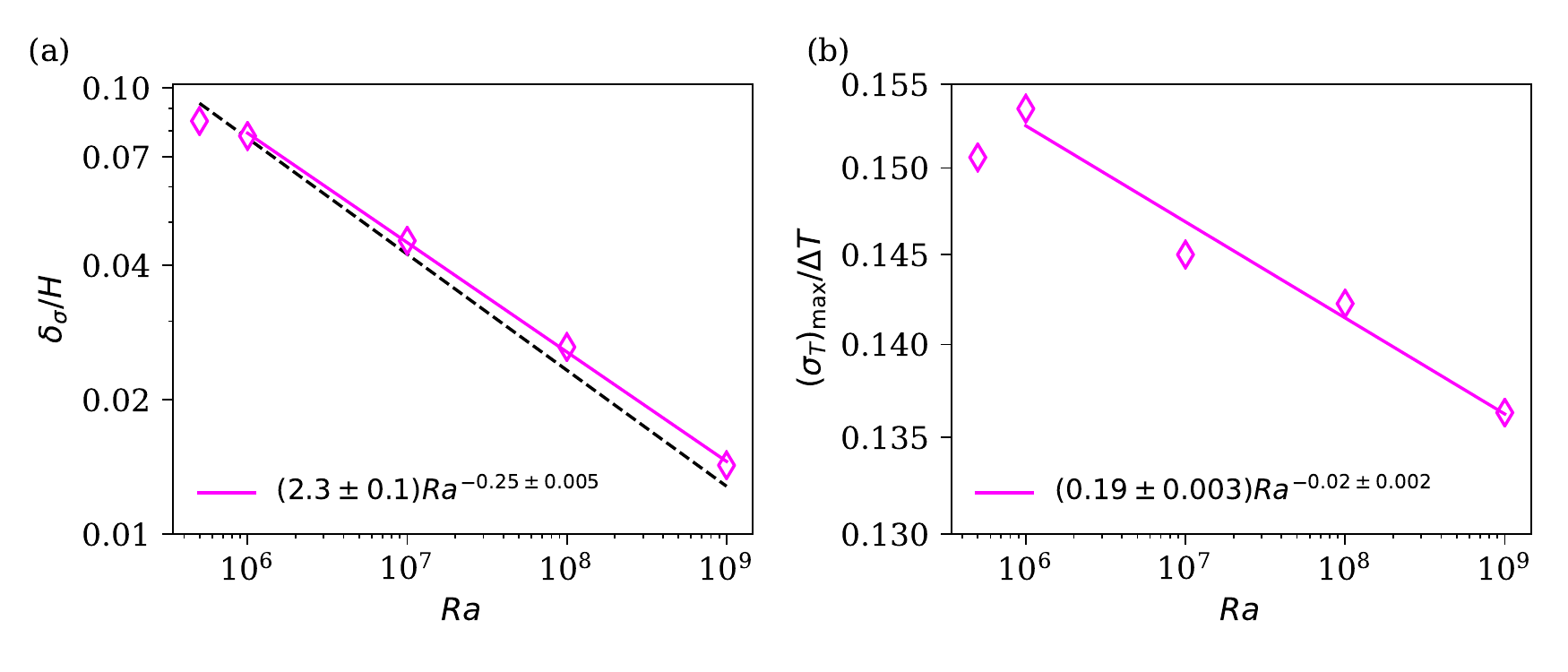}
}
\caption{(a) Thermal BL thickness extracted from the temperature variance profiles scales with $Ra$ similarly to the scaling of $\delta_{\langle T \rangle}(Ra)$, which is indicated by a black dashed line. (b) The maximum amplitude of the rms temperature fluctuations for $Ra \geq 10^6$ decreases very weakly with increasing $Ra$.}
\label{fig:delta_sigma}
\end{figure}
We plot  $\delta_\sigma$ as a function of $Ra$ in figure~\ref{fig:delta_sigma}(a) and observe that $\delta_\sigma$ decreases as a powerlaw except for the data point at $Ra = 5 \times 10^5$, which does not seem to follow the scaling very well. Weaker temperature fluctuations due to smaller $Ra$ might be the reason for this discrepancy. The best fit for $Ra \geq 10^6$ yields $\delta_\sigma/H = (2.3 \pm 0.1) Ra^{-0.25 \pm 0.005}$, which is close to the scaling of $\delta_{\langle T \rangle}$ (shown as a black dashed line in figure~\ref{fig:delta_sigma}(a)). The maximum amplitude of the temperature fluctuations is plotted as a function of $Ra$  in figure~\ref{fig:delta_sigma}(b), which again shows that the data at the lowest $Ra$ does not follow the trend for higher $Ra$.  The best fit for $Ra \geq 10^6$ yields $(\sigma_T)_\mathrm{max}/\Delta T = (0.19 \pm 0.003)Ra^{-0.02 \pm 0.002}$. Thus, $(\sigma_T)_\mathrm{max}/\Delta T$ decreases very weakly with increasing $Ra$ in our 2D RBC.

Note that $(\sigma_T)_\mathrm{max}$ indicates the disparity between the temperatures of the ambient fluid and the plumes at the edge of the thermal BL. It is observed in RBC that the approach of the horizontally- and temporally-averaged profile towards the bulk temperature becomes slower with increasing $Ra$ (see figure~\ref{fig:mean_BL_profile} for the mean thermal BL profiles in our flow)~\citep{Scheel:JFM2016, Shishkina:PRF2017}. Therefore, with increasing $Ra$, the temperature of the ambient fluid at the edge of the thermal BL becomes slightly closer to the temperature at the plate, which means that the contrast between the temperatures of the plumes and of the ambient fluid at the BL height decreases, as the temperature of the plumes remains nearly the same within the BL. This decreasing disparity at the edge of the BL with increasing $Ra$ yields a decreasing $(\sigma_T)_\mathrm{max}$ in figure~\ref{fig:delta_sigma}(b). Moreover, we find that $(\sigma_T)_\mathrm{max}/\Delta T \approx 0.14$ for $Ra = 10^7$ in our 2D RBC, which is larger than $(\sigma_T)_\mathrm{max}/\Delta T \approx 0.11$ observed by~\citet{Scheel:JFM2016} for $Pr = 0.021, Ra = 10^7$ in a cylindrical cell of aspect ratio one. This difference can be explained by the observation of \cite{Poel:JFM2013}, who noted that the thermal BL profile in 2D RBC is closer to the PBP profile compared to that in 3D RBC for similar parameters. We also observe this when we compare the mean thermal BL profile for $Ra = 10^7$ in our flow (see figure~\ref{fig:mean_BL_profile}) with the corresponding profile in figure~8 of \citet{Scheel:JFM2016}. Thus, the aforementioned temperature disparity at the edge of the thermal BL is smaller in 3D RBC, which results in a smaller $(\sigma_T)_\mathrm{max}$ in 3D compared to that in 2D RBC. A weakly decreasing $(\sigma_T)_\mathrm{max}$ in our 2D RBC indicates that the aforementioned temperature disparity at the thermal BL height decreases slowly or, in other words, the deviation of mean thermal BL profile from the PBP profile increases slowly in 2D RBC with increasing $Ra$.


\section{Thermal boundary layer profiles}		\label{sec:BL_pro}

We compare the temperature profiles in the BL region with the PBP profile, which is obtained by solving the following equations~\citep{Shishkina:NJP2010, Scheel:JFM2012}:
\begin{eqnarray}
\frac{\mathrm{d}^3 \Psi}{\mathrm{d}\xi^3} + \frac{1}{2} \Psi \frac{\mathrm{d}^2 \Psi}{\mathrm{d}\xi^2} & = & 0,  \label{eq:psi} \\
\frac{\mathrm{d}^2 \Theta}{\mathrm{d}\xi^2} + \frac{1}{2} Pr \Psi \frac{\mathrm{d} \Theta}{\mathrm{d}\xi} & = & 0 \label{eq:Theta}
\end{eqnarray}
with the boundary conditions
\begin{eqnarray}
\Psi(0) & = & 0, \hspace{5mm} \frac{\mathrm{d} \Psi}{\mathrm{d} \xi}(0) = 0, \hspace{5mm}  \frac{\mathrm{d} \Psi}{\mathrm{d} \xi}(\infty) = 1,  \label{eq:BC_psi} \\
\Theta(0) & = & 0, \hspace{5mm} \Theta(\infty) = 1. \label{eq:BC_Theta}
\end{eqnarray}
Here, $\Psi$ is the stream function, whose derivative yields the horizontal velocity, i.e., $u_x = \mathrm{d} \Psi / \mathrm{d}\xi$, and $\Theta$ is the normalized temperature defined as 
\begin{equation}
\Theta = \frac{T_\mathrm{bot}-T}{T_\mathrm{bot}-T_\infty}, \label{eq:scale}
\end{equation}
where $T_\mathrm{bot} = 1$ is the temperature at the bottom plate and $T_\infty$ is the temperature in the bulk region. The similarity variable $\xi$ is defined as $z/l$, with $l$ being the characteristic length scale of the thermal BL, which is $\delta_{\langle T \rangle}$ in the present case. We solve equations~(\ref{eq:psi}--\ref{eq:Theta}) together with the prescribed boundary conditions~(\ref{eq:BC_psi}--\ref{eq:BC_Theta}) using the shooting method to obtain $\Theta$ and $\Psi$ as a function of $\xi$.

We compare the temperature profiles measured at various horizontal positions at the plate with the PBP profile. However, we first compare the horizontally- and temporally-averaged profiles as they have been observed to show stronger deviations from the PBP profile~\citep{Shishkina:JFM2009}.


\subsection{Mean boundary layer profiles}

To compare our temperature profiles with the PBP profile, we transform $T(z)$ near the bottom plate to obtain $\Theta$ by using $T_\infty = T_\mathrm{bulk} = 0.5$ in  equation~(\ref{eq:scale}). We similarly transform $T(z)$ near the top plate as
$\Theta = (T-T_\mathrm{top})/(T_\infty-T_\mathrm{top})$, where $T_\mathrm{top} =
0$ is the temperature at the top plate.
\begin{figure}
\centerline{
\includegraphics[width=\textwidth]{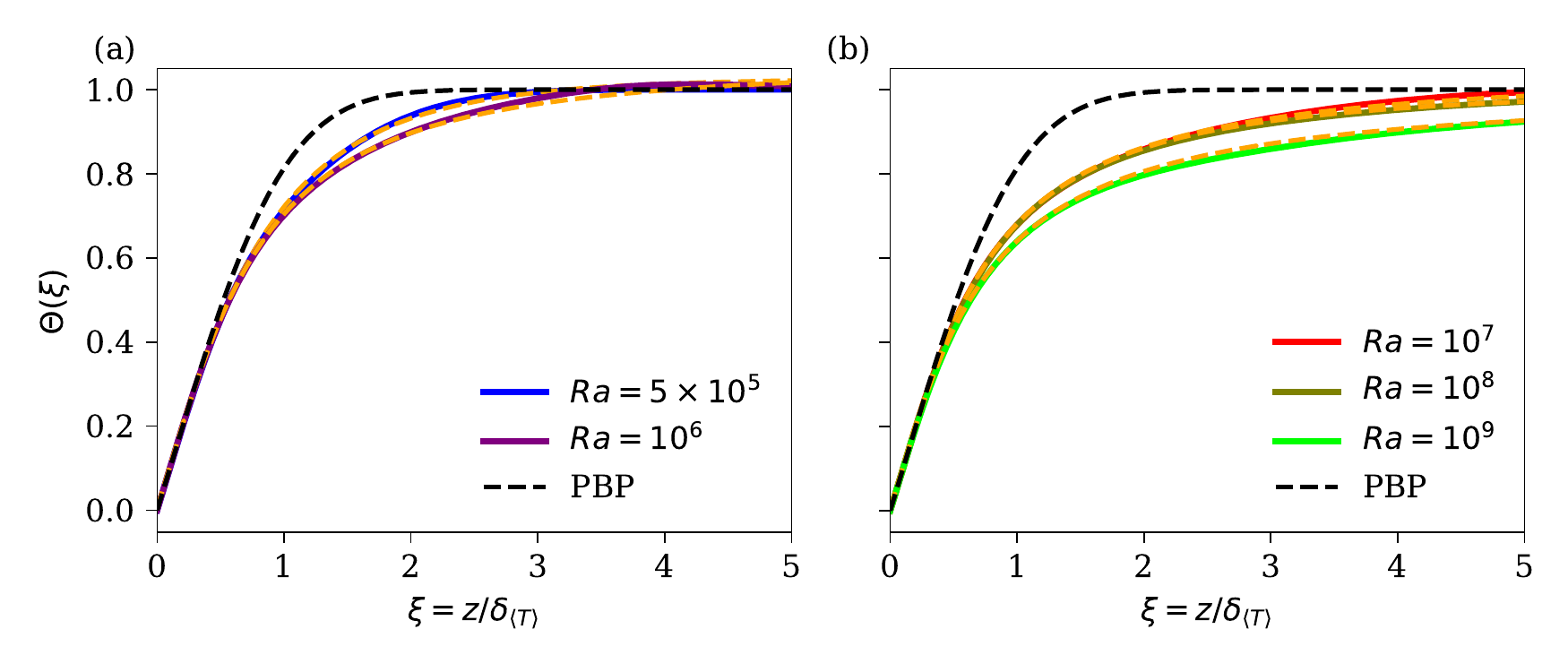}
}
\caption{BL profiles of the horizontally- and temporally-averaged temperature $\Theta$ deviate from the PBP profile (black dashed curve in both the panels) for all the Rayleigh numbers. However, all the profiles agree well with those computed from equation~(\ref{eq:Shishkina}) (orange dashed curves) with the coefficients provided in table~\ref{table:coeff}.}
\label{fig:mean_BL_profile}
\end{figure}
We compare the horizontally- and temporally-averaged profiles $\langle T \rangle_{x,t}$  near the plate by plotting $\Theta$ obtained by using $T = \langle T \rangle_{x,t}$ in equation~(\ref{eq:scale}) and in the aforementioned relation for $\Theta$ near the top plate as a function of $\xi =  z/\delta_{\langle T \rangle}$ in figure~\ref{fig:mean_BL_profile}. We further average over the top and bottom halves of the domain due to the top-bottom symmetry of our flow. We observe that the profiles for all the Rayleigh numbers deviate from the PBP profile, and the deviation increases with increasing $Ra$. Moreover, the approach to the bulk temperature becomes slower as $Ra$ increases. The reason for the deviation is that the RBC flow in a bounded domain does not satisfy the criteria for the PBP profile due to the presence of other effects, such as the emission of thermal plumes, buoyancy, pressure gradient, turbulent fluctuations, sidewalls, etc. Therefore, modified BL profiles in RBC have been suggested by incorporating these additional effects in the laminar BL equations~\citep{Shi:JFM2012, Shishkina:PRL2015, Shishkina:PRF2017,Ovsyannikov:EJMB2016, Ching:PRR2019}.
\citet{Shishkina:PRF2017, Ching:PRR2019} recently proposed a model of the horizontally- and temporally-averaged temperature profiles in the BL region by incorporating the effects of turbulent fluctuations in the laminar BL equations. They proposed that the temperature profile could be fitted with an equation of the form
\begin{equation}
\Theta(\xi) =  \frac{1}{b} \int_0^{b\xi} \left[ 1 + \frac{3 a^3}{b^3}(\eta - \arctan{(\eta)}) \right]^{-c} d\eta \, , \label{eq:Shishkina}
\end{equation}
where the coefficients $a, b, c$ can be obtained by fitting the  temperature profile with this equation. The equation~(\ref{eq:Shishkina}) was obtained by considering the variation of the turbulent diffusivity $\kappa_\mathrm{turb}$ near the plate. ~\citet{Shishkina:PRL2015} observed that $|\kappa_\mathrm{turb}/\kappa|$ can be approximated as $a_S^3\xi^3$ in the vicinity of the bottom plate, whereas  as $\xi$ in the logarithmic region far away from the plate.

\begin{figure}
\centerline{
\includegraphics[width=\textwidth]{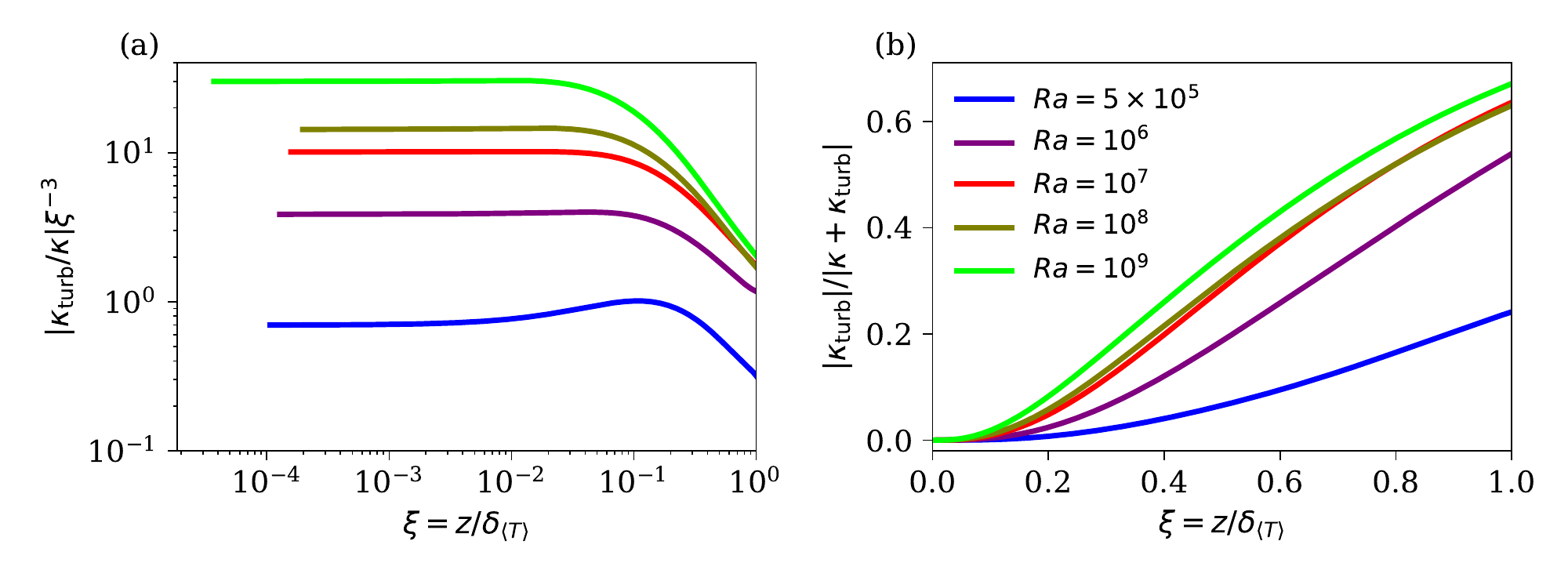}
}
\caption{(a) Vertical profiles of the horizontally-averaged turbulent diffusivity $|\kappa_\mathrm{turb}/\kappa|$ vary as $\xi^3$ in the vicinity of the bottom plate for all $Ra$, which is consistent with those observed in 3D RBC for different Prandtl numbers~\citep{Shishkina:PRL2015, Wang:JFM2018}. (b) Variation of $|\kappa_\mathrm{turb}|/|\kappa + \kappa_\mathrm{turb}|$ signifying the fraction of heat transport due to turbulent fluctuations indicates that the turbulent heat flux becomes increasingly important within the BL region with increasing $Ra$.}
\label{fig:kappa_t}
\end{figure}

\begin{table}
\begin{center}
\begin{tabular}{lccccc}
Run & $\mathrm{Ra}$ & $a$ & $b$ & $c$ & $a_S$ \\
1 & $5 \times 10^5$ & 0.92 &  3.77 & 9.47 & 0.94 \\
2 & $10^6$ & 1.32 & 2.75 & 2.64 & 1.58  \\
3 & $10^7$ & 1.47 & 3.06 & 2.41 & 2.16  \\
4 & $10^8$ & 1.42 & 3.29 & 2.86 & 2.43 \\
5 & $10^9$ & 1.79 & 3.85 & 2.22 & 3.07 \\
\end{tabular}
\caption{The coefficients $a,b,c$ are obtained from fitting the horizontally- and temporally-averaged profiles with equation~(\ref{eq:Shishkina}). These coefficients are used to compute the profiles exhibited as orange dashed curves in figure~\ref{fig:mean_BL_profile}. However, the coefficient $a_S$ is obtained from fitting $|\kappa_\mathrm{turb}/\kappa| = a_S^3 \xi^3$ in the region $\xi = 0$ to $0.05$ and differs from the coefficient $a$.}
\label{table:coeff}
\end{center}
\end{table}

Using our DNS data, we compute the turbulent diffusivity  defined as 
\begin{equation}
\kappa_\mathrm{turb}({\bm x}) = - \frac{\langle u_z^{\prime} T^{\prime} \rangle_t}{\p \langle T \rangle_t /\p z},
\end{equation}
where $u_z^{\prime}$ and $T^{\prime}$ are respectively the fluctuations in the vertical velocity and temperature from the time-averaged fields, which are defined as
\begin{eqnarray}
u_z({\bm x,t}) &=& \langle u_z \rangle_t ({\bm x}) + u_z^{\prime}({\bm x,t}), \\
T({\bm x, t}) &=& \langle T \rangle_t ({\bm x}) + T^{\prime}({\bm x,t}).
\end{eqnarray}
We plot the horizontally-averaged $|\kappa_\mathrm{turb}/\kappa|$ normalized by $\xi^3$ in figure~\ref{fig:kappa_t}(a) and observe that $\kappa_\mathrm{turb}$ indeed scales as $\xi^3$ near the plate for all the Rayleigh numbers. However, the $\xi^3$-scaling is satisfied only up to $\xi \approx 0.05$,  beyond which the scaling exponent decreases for all $Ra$. We obtain $a_S$  for all the Rayleigh numbers by fitting $|\kappa_\mathrm{turb}/\kappa| = a_S^3 \xi^3$ up to $\xi = 0.05$~\citep{Shishkina:PRL2015}. Figure~\ref{fig:kappa_t}(a) also shows that the prefactor $a_S$, which is a measure of the strength of turbulent fluctuations, increases with increasing $Ra$. We fit the temperature profiles shown in figure~\ref{fig:mean_BL_profile} with equation~(\ref{eq:Shishkina}) and obtain the coefficients $a,b,c$. We then compute the theoretical profiles  using equation~(\ref{eq:Shishkina}) with the obtained coefficients $a,b,c$, and show them in figure~\ref{fig:mean_BL_profile}. The figure exhibits that the temperature profiles obtained from our 2D DNS can be described well by equation~(\ref{eq:Shishkina}) with the proper choice of the coefficients $a,b,c$, which are summarized in table~\ref{table:coeff}.

\subsection{Local boundary layer profiles}

\begin{figure}
\centerline{
\includegraphics[width=0.9\textwidth]{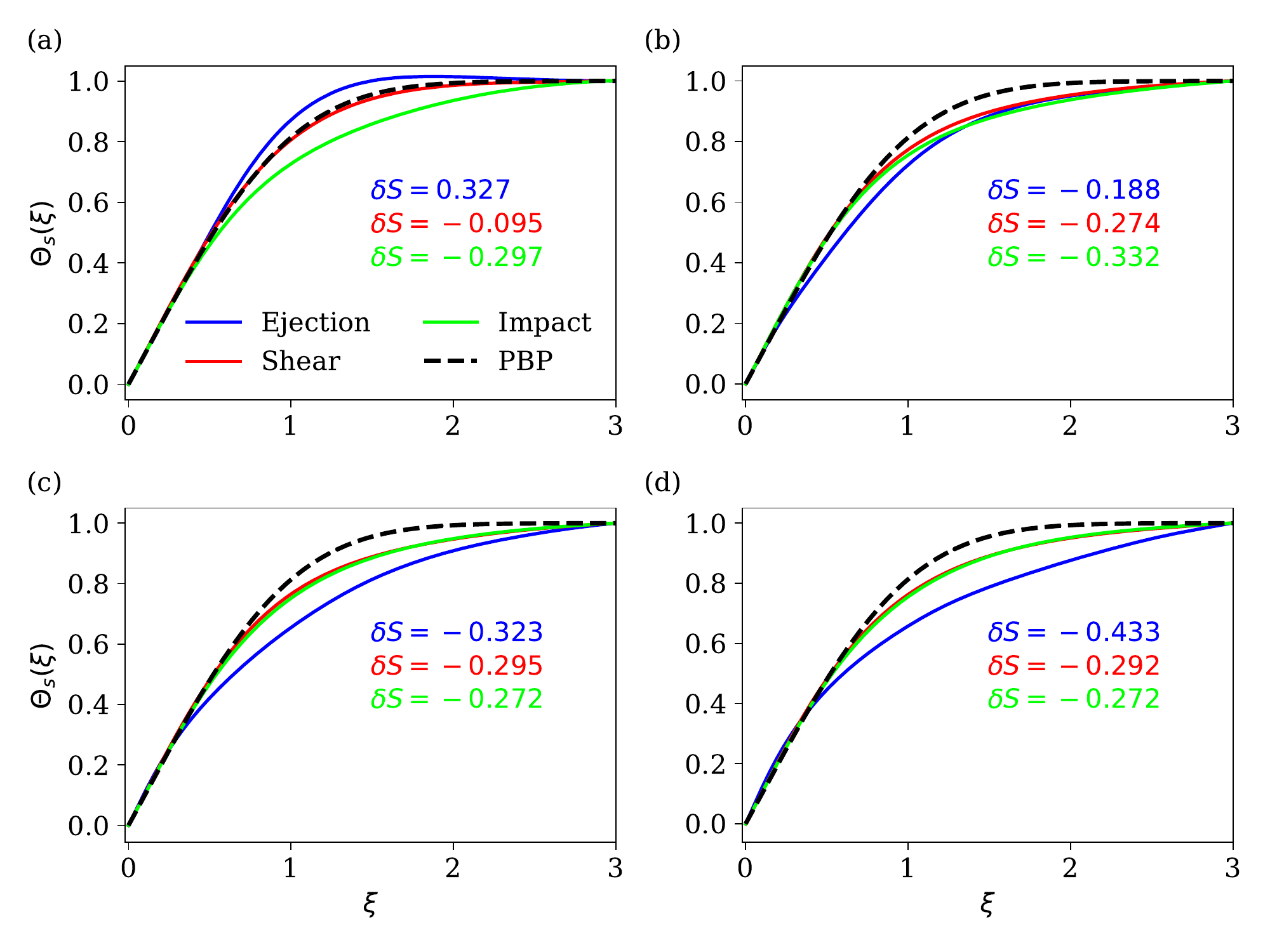}
}
\caption{BL profiles of the scaled-temperature $\Theta_s = \Theta(\xi)/\Theta(\xi = 3)$ in the ejection, shear, and impact regions for (a) $Ra = 10^6$, (b) $Ra = 10^7$, (c) $Ra = 10^8$, and (d) $Ra = 10^9$. Deviation from the PBP profile (indicated as a black dashed  curve in all the panels) can be observed in nearly all the profiles. The deviation $\delta S$ of the shape factor of the profiles from that of the PBP profile are also indicated with the corresponding colors. }
\label{fig:bl_local_profile}
\end{figure}

We now compare the local BL profiles in the ejection, impact, and shear regions with the PBP profile. To do this, we compute the time-averaged profiles $\langle T(x_0) \rangle_t$ at $x_0 = L/4, L/2, 3L/4$  and transform them using equation~(\ref{eq:scale}) to get the normalized temperature profile $\Theta(\xi)$. In figure~\ref{fig:bl_local_profile}, we plot the scaled-temperature $\Theta_s = \Theta(\xi)/\Theta(\xi = 3)$ as a function of $\xi = z/\delta_{\langle T \rangle}(x_0)$, i.e., $\xi$ is defined using the local BL thickness. We use this additional normalization because $\Theta$ does not saturate to unity in many of our profiles, and therefore, it may be normalized with a value of $\Theta$ in the region far from the BL, such as with $\Theta(\xi = 3)$~\citep{Stevens:PRE2012, Zhou:POF2011}. 

Figure~\ref{fig:bl_local_profile} exhibits that the scaled BL profiles $\Theta_s(\xi)$ agree with the PBP profile only in a region very close to the bottom plate, i.e., up to $\xi \approx 1/2$. For $Ra = 10^6$ (figure~\ref{fig:bl_local_profile}(a)), the profiles in the impact and ejection regions deviate from the PBP profile for $\xi \geqslant 0.5$. However, $\Theta_s(\xi)$ in the shear region agrees with the PBP profile  relatively well over the entire range of $\xi$. Moreover, $\Theta_s(\xi)$ in the ejection region for $Ra = 10^6$ exhibits overshoot, which is due to the growth of the corner roll in the impact region on the same plate. The overshoot in a profile indicates that the local temperature gradient is larger compared to that for the PBP profile. To investigate the reason for this overshoot, we look  at the instantaneous normalized profiles in the ejection region and find that not all the profiles exhibit overshoot. We observe that just before the occurrence of the overshoot the size of the corner roll in the impact region, i.e., near the opposite sidewall, grows. As a result, the impinging plumes are diverted towards the ejection region. This causes an increase of the local temperature gradient in the ejection region, which is reflected as overshoot in the corresponding instantaneous temperature profile. As the flow evolution is nearly periodic for $Ra = 10^6$ (see figure~\ref{fig:time_trace}(b) and supplementary movies), the aforementioned growth of the corner vortices occurs regularly, and thus, the overshoot is also present in the time-averaged profile. We do not observe the overshoot in the profiles for $Ra \geq 10^7$ as the corner flow structures become weaker for $Ra > 10^6$ in our low-$Pr$ RBC.
For $Ra \geq 10^7$, we find that $\Theta_s(\xi)$ in all the three regions deviate from the PBP profile. Additionally, $\Theta_s(\xi)$ in the shear and impact regions agree well with each other for the almost entire range of $\xi$ for $Ra \geq 10^7$, which is consistent with the observation from figure~\ref{fig:T_z_local}. We find that the profiles measured in the ejection region deviate most from the PBP profile.

The quality of agreement of the BL profiles with the PBP profile can be quantified by computing the shape factor of the profiles~\citep{Schlichting:book2004}, which is defined as
\begin{equation}
S = \delta_d/\delta_m, \label{eq:shape}
\end{equation}
where $\delta_d$ and $\delta_m$ are respectively the displacement and the momentum thicknesses of the profiles, and are computed as
\begin{eqnarray}
\delta_d & = & \int_0^\infty \left( 1 - \frac{\Theta(\xi)}{[\Theta(\xi)]_\mathrm{max}} \right) d\xi, \label{eq:dis_thick} \\
\delta_m & = & \int_0^\infty \left( 1 - \frac{\Theta(\xi)}{[\Theta(\xi)]_\mathrm{max}} \right) \left( \frac{\Theta(\xi)}{[\Theta(\xi)]_\mathrm{max}} \right) d\xi. \label{eq:mom_thick}
\end{eqnarray}
The shape factor of the PBP profile depends on the Prandtl number, and for $Pr = 0.021$ the shape factor $S_\mathrm{PBP} = 2.47$. Note that the shape factor of a profile indicates its tendency to quickly approach its asymptotic value; the larger the shape factor the faster the profile approaches its asymptotic value, and vice versa~\citep{Zhou:JFM2010, Scheel:JFM2012}. We compute the shape factors of the profiles shown in figure~\ref{fig:bl_local_profile} and indicate the deviation $\delta S = S - S_\mathrm{PBP}$ from the shape factor of the PBP profile with the corresponding colors in the same figure. Note however that, to compute $\delta_d$ and $\delta_m$ of the profiles, we perform integration only up to $\xi = 3$. We observe that $\delta S$ is negative for most of the profiles in figure~\ref{fig:bl_local_profile} except for $\Theta_s(\xi)$ in ejection region for $Ra = 10^6$, which exhibits the overshoot. We can see from figure~\ref{fig:bl_local_profile} that $\delta S$ for the profiles in the shear and impact regions do not differ much for $Ra \geq 10^7$, thus quantitatively indicating their similarity.

\begin{figure}
\centerline{
\includegraphics[width=0.9\textwidth]{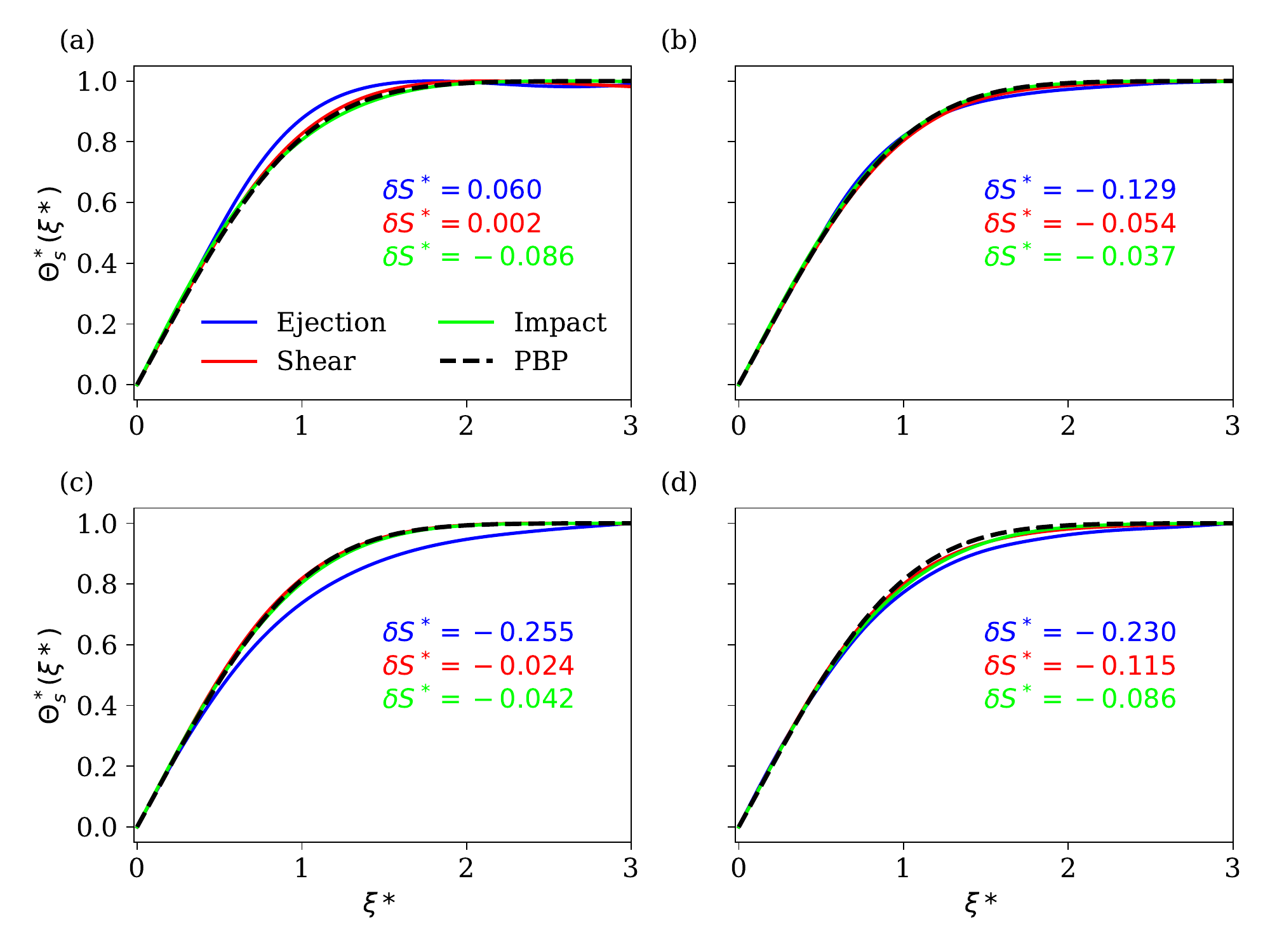}
}
\caption{Dynamically-rescaled thermal BL profiles for (a) $Ra = 10^6$, (b) $Ra = 10^7$, (c) $Ra = 10^8$, and (d) $Ra = 10^9$. The rescaled profiles in the shear and impact regions agree well with, whereas those in the ejection regions deviate from, the PBP profile (black dashed curve). The deviation of the shape factor $\delta S^*$ for all the profiles are also indicated with the corresponding colors.}
\label{fig:dyn_bl_profile}
\end{figure}

\citet{Zhou:PRL2010} observed that the BL profiles of the horizontal velocity in a high-$Pr$ RBC (water) agree better with the Prandtl-Blasius velocity profile if they are measured in a time-dependent frame of reference relative to the instantaneous BL width and then averaged in time. This dynamic rescaling was also applied to the thermal BL profiles for moderate and high Prandtl numbers and it was observed that the agreement of the rescaled profiles with the PBP profile becomes better~\citep{Zhou:JFM2010, Zhou:POF2011, Stevens:PRE2012, Shi:JFM2012, Scheel:JFM2012}. \citet{Zhou:POF2011} studied the structure of the BLs in a 2D square box for $Pr = 4.4$ and $Ra = 10^8$ and found that the dynamically-rescaled thermal BL profiles at nearly all the horizontal locations agree better with the PBP profile compared to the corresponding unscaled profiles.  Here, we want to test if this rescaling works for thermal BL profiles in a low-$Pr$ convection.

We thus construct a dynamically-varying frame of reference as
\begin{equation}
\xi^*(t) = z/\delta_T(x_0,t),
\end{equation}
and average the temperature profiles in this varying frame of reference as
\begin{equation}
\Theta^*(\xi^*) = \langle \Theta(t, z|z = \xi^*(t) \delta_T(x_0,t) \rangle_t.
\end{equation}
This enables us to average the temperature field at the same relative distances compared to the instantaneous BL thicknesses, which fluctuate strongly in our low-$Pr$ convection. We again compute the scaled profiles defined as $\Theta_s^*(\xi^*) = \Theta^*(\xi^*)/\Theta^*(\xi^*=3)$ and exhibit them in figure~\ref{fig:dyn_bl_profile}, which reveals that the dynamically-rescaled profiles in the shear and impact regions agree very well with the PBP profile for all the Rayleigh numbers. The profiles in the ejection region, however, deviate even after applying the dynamic rescaling, except for $Ra = 10^7$, where the agreement is rather good. 

We again compute the deviation $\delta S^* = S^*-S_\mathrm{PBP}$ for the rescaled profiles and indicate them in figure~\ref{fig:dyn_bl_profile}. We find for all the profiles that $|\delta S^*| < |\delta S|$, thus indicating that the agreement with the PBP profile becomes better if the profiles are measured in the dynamically-rescaled frame~\citep{Zhou:PRL2010}. Furthermore, $|\delta S^*|$ for the profiles in the ejection region increases with increasing $Ra$, which indicates that the deviations in the thermal BL profiles in this region become stronger as $Ra$ increases. Similar to $\delta S$, $\delta S^*$ for the profiles in the shear and impact regions are very similar (and closer to zero) for all the Rayleigh numbers. 

\subsection{Turbulent or not?}		\label{sec:log_pro}

\begin{figure}
\centerline{
\includegraphics[width=\textwidth]{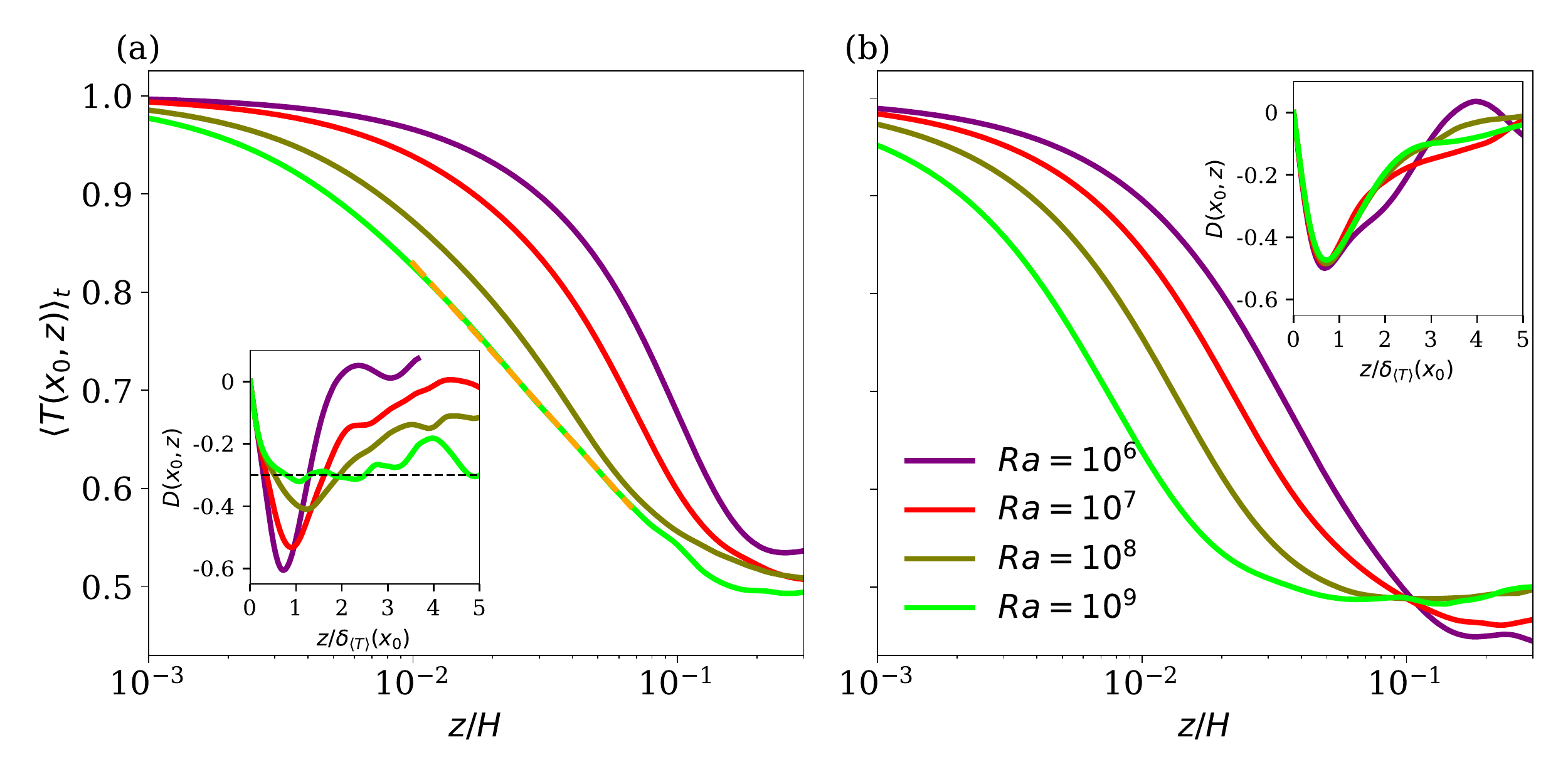}
}
\caption{Time-averaged temperature profiles in (a) the ejection region at $x_0 = L/4$ and (b) the impact region at $x_0 = 3L/4$ on a semilogarithmic scale. The profile for $Ra = 10^9$ in panel (a) exhibits a discernible logarithmic region and the resulting best fit curve is indicated as an orange dashed curve. Insets show the diagnostic function for the corresponding profiles as a function of the normalized vertical distance from the plate. A plateau region in $D(x_0,z)$ in panel (a)  for $Ra  = 10^9$ can be observed for $\delta_{\langle T \rangle} \lesssim z \lesssim 3 \delta_{\langle T \rangle}$.}
\label{fig:profile_log}
\end{figure}

We have observed in figure~\ref{fig:T_z_local} that the profiles  in the ejection region approach the bulk temperature slowly compared to those in the shear and impact regions. On the one hand, the rescaled profiles in the shear and impact regions are very similar to the PBP profile, which indicates that the fraction of the BL corresponding to these regions is laminar in the scalingwise sense. On the other hand, the profiles in the ejection region deviate conspicuously from the PBP profile, which implies that the local thermal BL properties in the ejection region differ from those of a laminar BL. Note that the shape factor of a turbulent BL profile is 1.28, which is smaller than $S_\mathrm{PBP}$ for $Pr = 0.021$. Thus, the increasing negative deviation in the shape factors of our profiles in the ejection region, which does not vanish even after applying the dynamic rescaling, indicates that the ejection region becomes increasingly turbulent with increasing $Ra$. As the turbulent BL profiles exhibit a logarithmic scaling~\citep{Ahlers:JFM2014, Poel:PRL2015, Schumacher:PRF2016, Zhu:PRL2018},  we plot the local profiles near the bottom plate on a semilogarithmic scale to explore the logarithmic behaviour of our profiles. Time-averaged profiles in the ejection region, i.e., at $x_0 = L/4$, are exhibited in figure~\ref{fig:profile_log}(a), where we observe in the profile for $Ra = 10^9$ that there exists a region between $z \approx 0.01$ and $z \approx 0.07$, which can be fitted as $\langle T(x_0, z) \rangle_t = A \log(z) + B$. The best fit yields $A = -0.30 \pm 0.001$ and $B = 0.23 \pm 0.001$, and the resulting best fit curve is shown as an orange dashed curve in figure~\ref{fig:profile_log}(a).  Figure~\ref{fig:profile_log}(b) shows the profiles in the impact region, i.e., at $x_0 = 3L/4$. We do not show the profiles in the shear region as they are very similar to the profiles in the impact region (see figure~\ref{fig:T_z_local}).  

Figure~\ref{fig:profile_log} reveals that the other profiles do not exhibit a discernible logarithmic region as the profile in the ejection region for $Ra  = 10^9$ does. 
The logarithmic behaviour of the profiles can be detected more clearly by looking at a diagnostic function $D(z) = dT/d \log z$, which should exhibit a plateau in the region, where the temperature profile exhibits a logarithmic scaling~\citep{Shishkina:JFM2009, Wagner:JFM2012, Zhou:JFM2013, Poel:PRL2015}. Insets of figure~\ref{fig:profile_log} show the diagnostic function $D(x_0,z) = \mathrm{d} \langle T(x_0) \rangle_t/\mathrm{d} \log z $ for the corresponding profiles, where we can see that no clear plateau can be observed in $D(x_0,z)$, except for $Ra = 10^9$ in the ejection region, where a plateau region exists for $\delta_{\langle T \rangle} \lesssim z \lesssim 3 \delta_{\langle T \rangle}$. This range roughly corresponds to the observed logarithmic range in the corresponding profile. We observe that $D(x_0,z)$ in figure~\ref{fig:profile_log}(a) in the region $\delta_{\langle T \rangle} \lesssim z \lesssim 3 \delta_{\langle T \rangle}$ is not a constant but fluctuates weakly around a mean value, which is due to a limited statistics available for this simulation. A longer simulation, and thus, a longer averaging would reduce the observed fluctuations in the plateau region.

Our observation that the temperature profile for $Ra = 10^9$ in the ejection region shows a discernible logarithmic range is similar to the observations of \citet{Poel:PRL2015}, who observed logarithmic temperature profiles in the ejection region in 2D RBC for $Pr = 1, Ra = 5 \times 10^{10}$. However, the slope $|A|$ of our profile is much larger compared to the slope of the logarithmic temperature profile in the bulk region observed by \citet{Ahlers:JFM2014} and \citet{Poel:PRL2015}. Moreover, the logarithmic region observed here overlaps with the buffer layer and differs from the logarithmic temperature profile in the bulk of the domain~\citep{Ahlers:JFM2014, Poel:PRL2015}.

Turbulent fluctuations in the BL region become stronger with increasing $Ra$ and fully turbulent BLs would prevail for sufficiently strong thermal forcing~\citep{Grossmann:JFM2000, Scheel:PRF2017}. However, for moderately strong thermal forcing the BLs are transitional, where a fraction of them is turbulent and the turbulent fraction grows continuously with increasing $Ra$~\citep{Scheel:JFM2016, Schumacher:PRF2016}. The strength of turbulent fluctuations within the BL region
can be investigated by looking at the turbulent fraction of the total heat flux, which we compute as~\citep{Wagner:JFM2012}
\begin{equation}
F_\mathrm{turb}({\bm x}) = \frac{\langle u_z^\prime T^\prime \rangle_t({\bm x})} {-\kappa \frac{ \p \langle T \rangle_t} {\p z}({\bm x}) + \langle u_z^\prime T^\prime \rangle_t({\bm x})}  = \frac{\kappa_\mathrm{turb}({\bm x})}{\kappa + \kappa_\mathrm{turb}({\bm x})}.
\end{equation}
We plot the horizontally-averaged turbulent fraction $F_\mathrm{turb}(z) = |\kappa_\mathrm{turb}(z)|/|\kappa + \kappa_\mathrm{turb}(z)|$ for all the Rayleigh numbers in figure~\ref{fig:kappa_t}(b), which reveals that $F_\mathrm{turb}(0) = 0$, as the heat flux is purely diffusive at the horizontal plates. As one moves from the plate towards the bulk region, turbulent fluctuations start to contribute to the total heat transport and their contribution increases with increasing distance from the plate. Moreover, at the same relative distance from the plate, $F_\mathrm{turb}(z)$ increases with increasing $Ra$. We observe that near $z \approx \delta_{\langle T \rangle}$ more than 50\% of heat is transported due to turbulent fluctuations for $Ra \geq 10^6$. Thus, the turbulent fluctuations are not negligible in the BL region in our low-$Pr$ RBC.
\begin{figure}
\centerline{
\includegraphics[width=\textwidth]{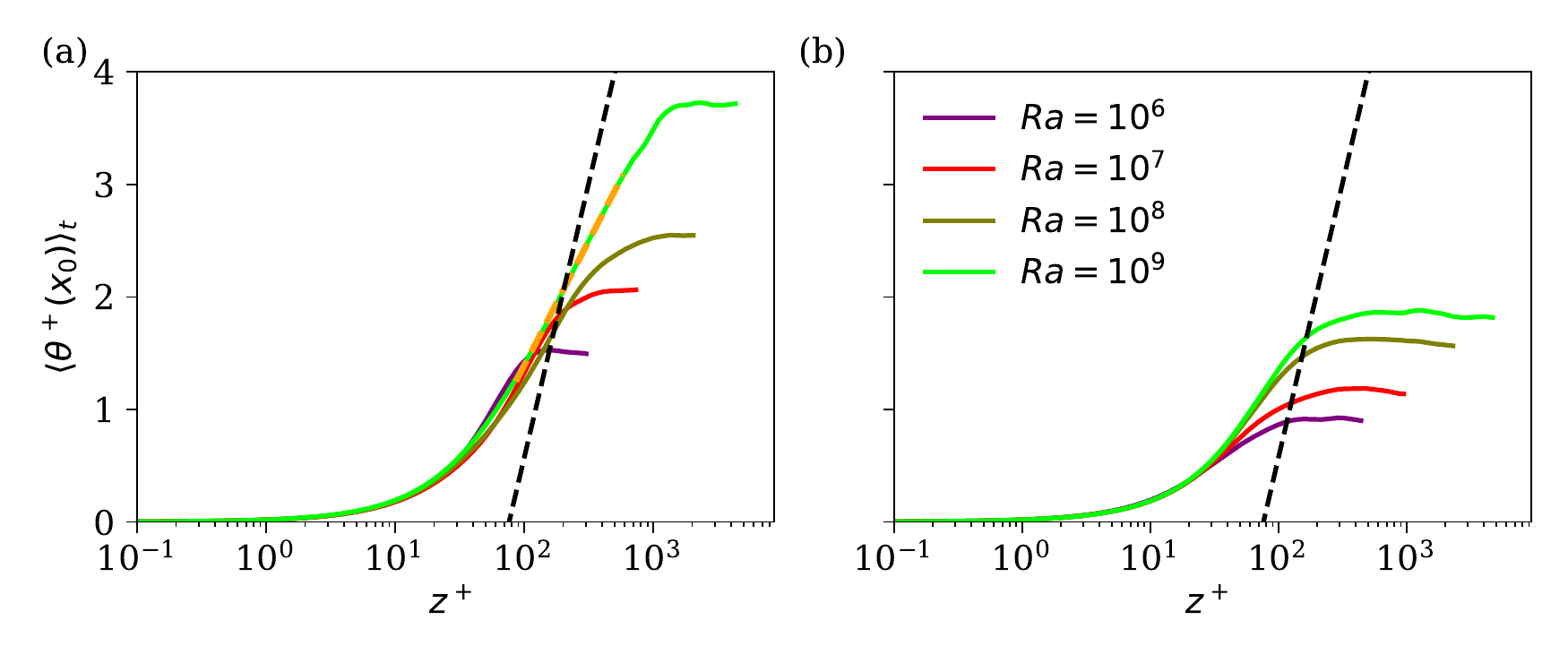}
}
\caption{Time-averaged temperature profiles in the inner wall units measured in (a) the ejection region at $x_0 = L/4$ and (b) in the impact region at $x_0 = 3L/4$. Similar to that in figure~\ref{fig:profile_log}(a), the profile for $Ra = 10^9$ in panel (a) exhibits a logarithmic scaling for $85 \leq z^+ \leq 580$ with a slope $\alpha_f \approx 0.96$, which is less than that of a fully turbulent BL (shown as a black dashed line in both the panels). This indicates that the BLs in our low-$Pr$ convection flow are not yet fully turbulent.}
\label{fig:profile_plus}
\end{figure}

Finally, we compare the local temperature profiles in our low-$Pr$ RBC with the fully turbulent thermal BL profile, which exhibits a logarithmic region in the overlap layer~\citep{Yaglom:ARFM1979}.
To do this, we plot the profiles in inner wall units~\citep{Schumacher:PRF2016, Scheel:PRF2017} by computing the local dimensionless  friction velocity and friction temperature as 
\begin{eqnarray}
u_\tau(x_0)  &=& \left( \frac{Pr}{Ra} \right)^{1/4} \left \langle \left[ \left. \left( \frac{\p u_x}{\p z} \right)^2 \right \vert_{x_0,z=0} \right]^{1/4} \right \rangle_t \, ,  \\
T_\tau(x_0) &=&  \left \langle \frac{1}{u_\tau(x_0,t) \sqrt{RaPr}}  \left. \frac{\p T}{\p z} \right \vert_{x_0,z=0} \right \rangle_t \, ,
\end{eqnarray}
where the derivatives at $x_0$  in the above equations  are averaged in the region $x_0-0.02L \leq x \leq x_0+0.02L$, with $x_0 = L/4, L/2, 3L/4$. The resulting local viscous length scale of the BL is given as 
\begin{equation}
z_\tau(x_0) = \sqrt{\frac{Pr}{Ra}} \, \frac{1}{u_\tau(x_0)}.
\end{equation}

To compare our profiles in the inner wall units, we define a temperature $\theta$ as
\begin{equation}
\theta(z) = (T_\mathrm{bot}-T(z))/\Delta T,
\end{equation}
and plot the rescaled-temperature $\langle \theta^+(x_0,z) \rangle_t = \langle \theta(x_0,z) \rangle_t /T_\tau(x_0)$ as a function of $z^+ = z/z_\tau(x_0)$ in figure~\ref{fig:profile_plus}. The rescaled profiles in the ejection region are exhibited in figure~\ref{fig:profile_plus}(a), where we observe that the profiles become more log-like with increasing $Ra$, and for $Ra = 10^9$, $\langle \theta^+(x_0,z) \rangle_t$ exhibits a discernible logarithmic scaling in some range of $z^+$. Following \citet{Yaglom:ARFM1979} and \citet{Kader:IJHMT1981}, the logarithmic temperature profile in the fully turbulent BL should scale as
\begin{equation}
\langle \theta^+(z) \rangle = \alpha \ln z^+ + \beta(Pr), \label{eq:KY}
\end{equation}
with $\alpha = 2.12$ and  
\begin{equation}
\beta(Pr) = (3.8 Pr^{1/3}-1)^2 - 1 + 2.12 \ln Pr.
\end{equation}
We fit the profile for $Ra = 10^9$ in the ejection region with equation~(\ref{eq:KY}) in the range of $z^+ = 85-580$, which corresponds to the fitting range shown in figure~\ref{fig:profile_log}(a). The best fit yields a slope $\alpha_f = 0.96 \pm 0.001$, which is less than $\alpha = 2.12$ for the fully turbulent thermal BL. As it is evident from figure~\ref{fig:profile_plus}, the other profiles do not exhibit a clearer logarithmic region, therefore we do not fit them with equation~(\ref{eq:KY}). Thus, our results indicate that the BLs in our 2D RBC for $Pr = 0.021$ are transitional and the highest $Ra$ achieved in this work is still not enough to yield a fully turbulent thermal BL. 

\section{Conclusions}	\label{sec:conclusion}

In this paper, we explored the structure of the thermal BL in low-$Pr$ RBC in a 2D square box by performing DNS for $Pr = 0.021$ and Rayleigh numbers  up to $10^9$, which was never achieved before. We interestingly found that the Nusselt numbers in our simulations agree reasonably well with those obtained by~\citet{Scheel:PRF2017} for $Pr = 0.021$ in a cylindrical cell. This similarity in $Nu$ implies that the scaling of the local thermal BL thickness as well as its horizontal structure observed in our 2D RBC might also be similar in 3D convection for low Prandtl numbers. The LSC yields three distinct flow regions at the horizontal plates, and we found that the properties of the thermal BL are different in these regions. The temperature profiles measured in the plume-ejection region approach the bulk temperature slowly compared to the profiles in the shear and impact regions. We observed that the thermal BL profiles in all regions deviate from the PBP profile and the strongest deviations are found in the ejection region. This is because the turbulent fluctuations are stronger in the ejection region, and therefore, the local BL properties in this region deviate the most from the properties of a laminar BL. The dynamically-rescaled profiles~\citep{Zhou:PRL2010} in the shear and impact regions agree well with the PBP profile for all the Rayleigh numbers, suggesting that these regions in the BL are laminar in the scalingwise sense. The rescaled profiles in the ejection region, however, exhibit persistent deviations~\citep{Shi:JFM2012} for all the Rayleigh numbers in our study.  By comparing our profiles with the turbulent BL profile, we concluded that the thermal BLs in our low-$Pr$ convection are transitional and become increasingly turbulent with increasing $Ra$~\citep{Schumacher:PRF2016}.

The horizontal structure of the thermal BL in low-$Pr$ convection has not been investigated earlier, and therefore to do this, we computed the time-averaged local BL thicknesses $\delta_{\langle T \rangle}(x)$ at the top and bottom plates. Our findings revealed that the $\delta_{\langle T \rangle}(x)$ are larger in the ejection region and decrease as the impact region is approached. Thus, the local thermal BL thickness in our 2D RBC grows in the downstream direction. However, $\delta_{\langle T \rangle}(x)$ in our flow neither grows as $\sqrt{x}$ as in a laminar BL nor as $x$ as in a turbulent BL~\citep{Schlichting:book2004}. We moreover found that $\delta_{\langle T \rangle}(x)$  at every horizontal location decreases as $Ra^{-\beta(x)}$, with the exponent $\beta(x)$ depending on the position at the plates. We found that the local exponents $\beta(x)$ in the ejection region are larger compared to those in the impact and shear regions. As $\delta_{\langle T \rangle}(x)$ is inversely proportional to the diffusive heat flux at the plate, we found that the local diffusive heat flux at the plate is smaller in the ejection region compared to those in the shear and impact regions. Moreover, the diffusive flux at the plate generally increases in the ejection region, whereas decreases in the other two regions with increasing $Ra$, which implies that the local BL thicknesses scale differently with $Ra$ compared to the scaling of mean BL thickness. A position-dependent exponent $\beta(x)$ implies that the horizontal variation of the BL thickness becomes weaker with increasing $Ra$, which is  due to the growing strength of LSC with increasing $Ra$ in our flow. We estimated that $\delta_{\langle T \rangle}(x)$ might be the same throughout the plate for $Ra \geq 8 \times 10^{12}$ in our low-$Pr$ convection, provided that the observed scalings of the local BL thicknesses hold also for the larger $Ra$.

Our observation of the position-dependent properties of the temperature field near the isothermal plates in low-$Pr$ convection is consistent with a similar picture found in high- and moderate-$Pr$ convection~\citep{He:PRL2019, Poel:PRL2015, Zhu:PRL2018}. However, as the temporal evolution of the thermal BL properties is correlated with those of the viscous BL~\citep{Zhou:JFM2010, Shi:JFM2012}, it is crucial to study the properties of viscous BLs in our low-$Pr$ RBC for its improved understanding. This is currently underway and will be reported elsewhere.

\section*{Acknowledgements}
The author thanks Katepalli R. Sreenivasan, J\"org Schumacher, Mahendra K. Verma, and Janet D. Scheel for helpful comments on the manuscript. The author is also grateful to the anonymous reviewers for their fruitful suggestions. This research was carried out on the High Performance Computing resources at New York University Abu Dhabi, as well as on {\sc Makalu} cluster at Technische Universit\"at Ilmenau, Germany. This work was supported by NYUAD Institute Grant G1502 ``NYUAD Center for Space Science."

\section*{Declaration of interests}
The author reports no conflict of interest.

\section*{Supplementary movies}
Supplementary movies are available at https://doi.org/10.1017/jfm.2020.961.

\section*{Author ORCID} A. Pandey, https://orcid.org/0000-0001-8232-6626

\end{document}